\documentclass[11pt]{article}

\usepackage{amsmath,amssymb,amsthm}
\usepackage[dvips]{graphicx}
\usepackage{psfrag}
\usepackage{fullpage}

\newcommand{\Ref}[1]{(\ref{#1})}

\def \sss{\subsubsection}

\newcommand{\Z}{\mathbb{Z}}

\newcommand{\R}{\mathbb{R}}
\newcommand{\C}{\mathbb{C}}
\DeclareMathOperator{\tr}{tr}

\def\be{\begin{equation}}
\def\ee{\end{equation}}
\def\bes{\begin{eqnarray}}
\def\ees{\end{eqnarray}}
\def\nn{\nonumber}
\def\arr{\rightarrow}
\def\harr{\hookrightarrow}

\def\la{\langle}
\def\ra{\rangle}
\def\f{\frac}
\def\tl{\tilde}
\def\wtl{\widetilde}

\def\what{\widehat}

\def \pp{\partial}
\def\eb{{\bar{e}}}
\def\fb{{\bar{f}}}
\def\vb{{\bar{v}}}
\def\wb{{\bar{w}}}
\def\e{\epsilon}
\def\S{{\cal S}}

\def\bt{\bar{\theta}}
\def\bra{\langle}
\def\ket{\rangle}
\def\zz{{\cal Z}}
\def\ii{{\cal I}}
\def \v{\vec}

\newcommand{\lalg}[1]{\mathfrak{#1}}  
\newcommand{\SU}{\mathrm{SU}}
\newcommand{\SL}{\mathrm{SL}}
\newcommand{\SO}{\mathrm{SO}}
\newcommand{\U}{\mathrm{U}}
\newcommand{\su}{\lalg{su}}

\newcommand{\mat}[2]{\left(\begin{array}{#1} #2\end{array}\right)}

\def \nn{\nonumber}

\title{Ponzano-Regge model revisited III:\\
 Feynman diagrams and Effective field theory}

\author{Laurent Freidel$^{1,2}$\thanks{email: lfreidel@perimeterinstitute.ca},
 Etera R. Livine$^1$\thanks{email: elivine@perimeterinstitute.ca} \\
\\ \centerline{\footnotesize \it $^1$Perimeter Institute for
Theoretical Physics} \\
\centerline{\footnotesize \it 35 King st N, Waterloo, ON, Canada
N2J 2W9. }\\
\centerline{\footnotesize \it $^2$Laboratoire de Physique, \'Ecole Normale Sup{\'e}rieure de Lyon} \\
\centerline{\footnotesize \it 46 all{\'e}e d'Italie, 69364 Lyon
Cedex 07, France.} }
\date{}

\begin{document}

\maketitle




\begin{abstract}
We study the no gravity limit $G_{N}\arr 0$ of the Ponzano-Regge amplitudes  with massive
particles and show that we recover in this limit Feynman graph amplitudes
(with Hadamard propagator) expressed as an abelian spin foam model.
We show how the $G_{N}$ expansion of the Ponzano-Regge amplitudes can be resummed.
This leads to the conclusion that the effective dynamics of quantum particles coupled
to quantum 3d gravity can be expressed in terms of
an effective new non commutative field theory which respects the principles of doubly special relativity.
We discuss the construction of Lorentzian spin foam models including Feynman propagators.

\end{abstract}

\maketitle

\tableofcontents

\newpage

\section{Introduction}

Spin Foam models offer a  rigorous framework implementing a path
integral  for quantum gravity \cite{reviews}. They provide a
definition of a quantum spacetime in purely algebraic and
combinatorial terms and describe it as generalized two-dimensional
Feynman diagrams with degrees of freedom propagating along
surfaces. Since these models were introduced, the most pressing
issue has been to understand their semi-classical limit, in order
to check whether we effectively recover general relativity and
quantum field theory as low energy regimes and in order to make
physical and experimental predictions carrying a quantum gravity
signature. A necessary ingredient of such an analysis is the
inclusion of matter and particles in a setting which has been
primarily constructed for pure gravity. On one hand, matter
degrees of freedom allow to write physically relevant
diffeomorphism invariant observables, which are needed to fully
build and interpret the theory. On the other hand, ultimately, we
would like to derive an effective theory describing the
propagation of matter within a quantum geometry and extract
quantum gravity corrections to scattering amplitudes and
cross-sections.

In the present work, we study  these issues in the case of  a
three-dimensional spacetime. Three-dimensional gravity is seen as
a toy model in the search of a consistent full quantum general
relativity. It is known to be an integrable system carrying a
finite number of degrees of freedom. Nevertheless, despite its
apparent simplicity, we face many of the same mathematical and
interpretational problems than when addressing the quantization of
four-dimensional gravity. In this context, the Ponzano-Regge model
was actually the first model of quantum gravity ever written
\cite{PonzanoRegge}. It gives an explicit prescription for the
scalar product of Euclidean three-dimensional gravity as a state
sum over discretized geometries. As the simplest non-trivial spin
foam model, it is the natural arena to investigate the particle
couplings to quantum geometry and the semi-classical regime of the
theory. Recently, this model has been studied in much detail. The
paper \cite{PR} has tackled particle insertions and shown how they
can be understood as a partial gauge fixing of the state sum
model. As a result, explicit  quantum amplitudes for massive and
spinning particles coupled to gravity were constructed. The work
in \cite{PRII} established an explicit link between the
Ponzano-Regge quantum gravity and the traditional Chern-Simons
quantization, and identified a $\kappa$-deformation of the
Poincar\'e group as the relevant symmetry group of spin foam
amplitudes. In this article we follow the same line of thoughts
focusing on the relation between spin foam amplitudes  and usual
Feynman graph evaluations.

The first step is to show that  the usual Feynman graph
evaluations (with the Hadamard propagator) for quantum field
theory on a three-dimensional flat Euclidean spacetime can be
identified through a duality transform as spin foam amplitudes of
an abelian model. This construction can be extended to the
non-abelian case, in which case we describe Feynman diagrams for
particles propagating in 3d quantum gravity. The natural mass unit
for particles is the Planck defined solely from the Newton
constant for gravitation $m_P=1/\kappa=1/4\pi G$.

We define a gravity-less limit of the full quantum gravity  theory
as the $G\arr 0$ limit. And we show that the Ponzano-Regge
amplitudes constructed in \cite{PR} reduce to the amplitudes of
the abelian model with the standard Feynman evaluations: as
required, the classical limit of matter insertion is the standard
quantum field theory.

Then, at $G\ne 0$, the spin foam amplitudes are to be  interpreted
as providing the Feynman graph evaluation of particles coupled to
quantum gravity. We study the perturbative $G$ expansion of the
spin foam amplitudes. Remarkably, this expansion can be re-summed
and expressed as the Feynman graphs of a non-commutative braided
quantum field theory with deformation parameter $G$, which thus
describes the effective theory for matter in quantum gravity. Any
deformed Poincar\'e theory usually suffers from a huge ambiguity
\cite{Jerzy} coming from what should be identify as the physical
energy and momenta since the introduction of the Planck scale
allows non-linear redefinitions. This ambiguity can also be
understood as an ambiguity in the identification of the
non-commutative space-time. Our work show that the Ponzano-Regge
model naturally defines a star product and a duality between space
and momenta, therefore no ambiguity remains once we identify
quantum gravity as being responsible for the effective deformation
of the Poincar\'e symmetry.

This realizes explicitly, for the first time from first
principles, the now popular idea that quantum gravity will
eventually lead to an effective non-commutative field theory
incorporating the principle of Doubly special relativity
\cite{NCgeometry}.

When we also take into account  a non vanishing cosmological
constant $\Lambda$, 3d Euclidean quantum gravity is described by
the Turaev-Viro model whose coupling to matter was investigated
by Barrett et al. \cite{Barrett1,Barrett2}. It is based on the quantum group
$\su_q(2)$. $\Lambda$ and $G$ now define natural length and mass
units, bounding the physical quantities by above and below. We
define different classical limits, which correspond to sending
some of these units to zero or infinity. Interestingly, they all
correspond to the same limit of the deformation parameter  $q\arr
1$. However they define different limits of the Feynman graph
evaluations, depending on which units we use to describe the mass
and length of physical objects. We define
 a hyperbolic spin foam models corresponding to a
space-time at $\Lambda\leq 0$. We study  the
gravity-less limit $G\arr 0$ of the Turaev-Viro model and of
the hyperbolic state sum model.

Finally, we tackle the issue  of causality of the particle
amplitudes. Indeed, in the Euclidean framework, we have dealt up
to now with the Hadamard propagator. It is non-oriented
in time and non-causal. In the Lorentzian spin foam model, it is
possible to discuss the different propagators -Wightman, Feynman
and Hadamard- and write down the corresponding amplitudes in the
non-abelian set-up. This means that we now have the explicit
proposal for the amplitudes of the non-commutative quantum field
theory describing the effective theory of 3d Lorentzian quantum
gravity.

\section{Spin Foams as the topological dual of Feynman diagrams: the abelian case}
\label{abelian}

We consider the following Feynman diagram amplitudes associated to an oriented graph $\Gamma$ for
a massive relativistic scalar field in the 3d Euclidean spacetime:
\be
I_\Gamma=\int\prod_v d\vec{x}_v\,\prod_e G_{m_e}(\vec{x}_{t(e)}-\vec{x}_{s(e)}),
\ee
where $v$ and $e$ label the vertices and the edges of the graph $\Gamma$, $m_e$ the mass
of the particles living on the edge $e$ and $G_m$ the propagator of the theory.
The $\vec{x}$ correspond to the positions of the particles/vertices of the graph
(they are 3-vectors).
Here we take:
\be
G_m(\vec{x})=\f{\sin m|x|}{|x|}.
\label{usual}
\ee
Let us insist on the fact that we are using the  Hadamard  function, and not the Feynman propagator.
So we should call {\it Feynman-Hadamard evaluation}, but we keep it to Feynman evaluation whenever
there can not be any confusion. We will later discuss the use and the implementation of the Feynman
propagator in the spinfoam framework, when dealing with the Lorentzian version of the models.
In the present section, we show how we can recast the Feynman evaluation as a spinfoam amplitude
through a duality transform.

Let us start by switching to the Fourier transform.  For this purpose we use the Kirillov formula:
\be
G_m(\vec{x})=m\int_{{\cal S}^2}d^2\vec{n} \, e^{im\vec{x}\cdot\vec{n}},
\label{kiri}
\ee
then we change the space of integration from ${\cal S}^2$ to $\R^3$ using:
\be
\int_{{\cal S}^2}d^2\vec{n}=
\f{1}{4\pi m^2}\int_{\R^3}d^3\vec{p}\,\delta(|p|-m).
\ee
This leads to:
\be
I_\Gamma=\int\prod_v d\vec{x}_v\,\int\prod_ed^3\vec{p_e}\,
\f{\delta(|p_e|-m_e)}{4\pi m_e}\prod_ee^{i\vec{p_e}\cdot(\vec{x}_{t(e)}-\vec{x}_{s(e)})}.
\label{abevaluation}
\ee
Integrating over the $\vec{x}_v$, we finally get:
\be
I_\Gamma=\int\prod_ed^3\vec{p_e}\,\f{\delta(|p_e|-m_e)}{4\pi m_e}
\prod_v\delta\left(\sum_{e|v=t(e)}\vec{p}_e-\sum_{e|v=s(e)}\vec{p}_e\right).
\label{evaluation}
\ee
The $\delta(|p|-m)$ fixes the norm of the momenta to be the mass\footnote{$\delta(|\vec{p}|-m)/2m$ is equivalent to the usual $\delta(|\vec{p}|^2-m^2)$.} and
the $\delta(\sum \vec{p}_e)$ at the vertices express momentum conservation.

\medskip

Let us now go to the topological dual and express this amplitude
as a Feynman evaluation on the dual graph.
A similar duality transformation on Feynman graphs was performed in
\cite{Emil}.

More precisely, let us
embed the graph $\Gamma$ in a surface $S$. This amounts to
defining faces $f$ (as sequences of edges) on the graph, each edge
belonging to two faces. Then one can define the topological dual,
the dual vertices $\vb$ being the faces, the dual edges $\eb$
linking the $\vb$ being associated to the original edges $e$ and
the dual faces corresponding to the initial vertices $v$.

We would like to solve the constraint on the dual faces  $\fb$ (i.e at the vertices $v$)
imposed by the $\delta(\sum_{\eb\in\pp \fb}\e_{\fb}(\eb)\vec{p}_{\eb})$ (where
$\e_{\fb}(\eb)$ is a sign recording the orientation of the edge
$\eb$ relatively to the face $\fb$).
From the algebraic topology point of view, $p$ is a field valued on the edges $\eb$, so it is
a 1-form. Its derivate $\pp p$ is a 2-form ie a field valued on faces $\fb$. More precisely:
$$
(\pp p)_{\fb}=\la \pp \fb,p\ra =\sum_{\eb\in\pp \fb}\e_{\fb}(\eb)\vec{p}_{\eb}.
$$
Therefore we are imposing that $\pp p=0$. If the (co)homology group $H^1(S)=H_1(S)$ is trivial,
all closed 1-forms are exact. Then there exists a field $\vec{u}$ valued on vertices
$\vb$ such that $p=\pp u$ i.e $p_{\eb}=u_{t(\eb)}-u_{s(\eb)}$,
whith $t(\eb), s(\eb)$ being the starting and terminal vertices of $eb$. More generally $H^1$ is
generated by the cycles of $S$. So if $S\sim{\cal S}^2$, everything is straightforward. But if the genus
of $S$ is $g\ge 1$, then we also need variables $\vec{u}_C$ attached to the cycles of $S$.
These cycles $C$ are defined, similarly to faces, as sequences of edges $\eb$. And now the solution
to $\pp X$ are given by:
\be
p_{\eb}=u_{t(\eb)}-u_{s(\eb)}+\sum_{C\ni \eb}\e_C(\eb)u_C,
\ee
where $\e_C(\eb)=\pm$ depending on the orientation of the dual edge $\eb$ along the cycle $C$.

A simple counting argument could reproduce the same output. Indeed we have $E$ (number of edges)
variables $p_{\eb}$ and $V-1$ ($V$ number of vertices) constraints. So we would need
$E-(V-1)=F-\chi+1$ variables to parametrize the solutions to this linear system, with
$\chi$ the Euler characteristic of the surface $S$. As we use variables associated to the
faces (i.e the dual vertices), let us allow for a closure relation on these face variables, so
we would need $F-\chi+2$ variables. If $S$ is orientable then, this gives $F+2g$ variables:
one variable per face and one variable per cycle. The same holds for non-orientable
surfaces with $F+g$ variables.

Finally the Feynman graph evaluation reads:
\be
I_\Gamma=\int_{\R^3}\prod_{\vb}d\vec{u}_{\vb}\prod_{C\,{\mathrm cycles}}d\vec{u}_C\,
\prod_e\f{\delta(|\vec{p}_{\eb}|-m_e)}{4\pi m_e}
\qquad\textrm{with}\qquad
p_{\eb}=u_{t(\eb)}-u_{s(\eb)}+\sum_{C\ni \eb}\e_C(\eb)u_C.
\ee
The reader can find examples for $S={\cal S}^2$ and $S=T^2$ in appendix.

More generally, let us consider a triangulated surface $S$ and the graph $\Gamma$ embedded in the triangulation (but not necessarily covering the whole triangulation). We could more generally consider a generic cellular decomposition of $S$ without loss of generality.  We attach variables to every face/triangle, or equivalently to all dual vertices $\vb$, and to every cycle of $S$. Then the Feynman evaluation simply reads:
\be
I_\Gamma=\int_{\R^3}\prod_{\vb\in S}d\vec{u}_{\vb}
\prod_{C\textrm{ cycles of } S}d\vec{u}_C\,
\prod_{e\in\Gamma}\f{\delta\left(|\vec{u}_{t(\eb)}-\vec{u}_{s(\eb)}+
\sum_{C\ni \eb}\e_C(\eb)\vec{u}_C|-m_e\right)}{4\pi m_e}
\prod_{e\in S\setminus\Gamma}\delta(\vec{u}_{t(\eb)}-\vec{u}_{s(\eb)}),
\label{2d}
\ee
where the $\delta$ functions on the edges in $S\setminus \Gamma$ allows to 'collapse' the triangulation of $S$ down to the graph $\Gamma$.

\medskip

We now embed the Feynman diagram in the three-dimensional spacetime and recast the Feynman evaluation in terms of the three-dimensional objects. Let us start with a three-dimensional triangulation $\Delta$
(or more generally a cellular decomposition), on which the graph $\Gamma$ is drawn. We now would like to express the Feynman amplitude in terms of variables living on the structures (vertices, edges, faces or 3-cells) of $\Delta$ of its dual. We remind that the (2-skeleton of the) topological dual of $\Delta$ is called the {\it spin foam}.

Let us choose a surface $S\subset \Delta$ in which $\Gamma$ is faithfully embedded
\footnote{We require that the components of $S$ when the graph $\Gamma$ is removed are all
homeomorphic to a  2d disk.}
 and rewrite the formula \Ref{2d}. $S$ is a collection of faces $f$ (triangles more exactly if $\Delta$ is strictly speaking a triangulation) to which we associate vectors $\vec{u}_f$. Equivalently, one can consider the dual edges $e^*$ transverse to the  faces $f$ and the corresponding variables $\vec{u}_{e^*}$. Then one can directly re-write the Feynman evaluation as:
\be
I_\Gamma=\int_{\R^3}\prod_{f\in S}d\vec{u}_{f}
\prod_{C\textrm{ cycles of } S}d\vec{u}_C\,
\prod_{e\in\Gamma}\f{\delta\left(|\vec{u}_{t(e)}-\vec{u}_{s(e)}+
\sum_{C\ni e}s_C(e)\vec{u}_C|-m_e\right)}{4\pi m_e}
\prod_{e\in S\setminus\Gamma}\delta(\vec{u}_{t(e)}-\vec{u}_{s(e)}),
\label{ig}
\ee
where $t(e)$ and $s(e)$ are the two {\it faces} of $S$ adjacent to the edge $e\in S$, and where $e\in C$ means that the link $\eb$ dual to the edge $e$ on the surface $S$ belongs to the cycle $C$.
We give an example for $S={\cal S}^2, \Delta={\cal S}^3$ in appendix.

\medskip

Let us now show how such an amplitude arises from spin foam models. Once the three-dimensional triangulation $\Delta$ is chosen, one can define the partition function of the (topological) spin foam model corresponding to the quantization of an abelian BF theory for the gauge group $\R^3$ on the dual of $\Delta$. One associate group variables -here vectors- $\vec{u}_{e^*}$ to every dual oriented edge $e^*$ (or equivalently to every face of $\Delta$) and one requires that the holonomies around all dual faces $f^*$ are trivial:
\be
Z^{(ab)}_\Delta=
\int_{\R^3}\prod_{e^*}d\vec{u}_{e^*}\prod_{f^*}
\delta(\vec{u}_{f^*}),
\ee
where we have introduced the notation
$$
\vec{u}_{f^*}\equiv\sum_{e^*\in \pp f^*}\e_{f^*}(e^*)\vec{u}_{e^*}.
$$
In order to introduce particles, it is more convenient to expand the $\delta$ function in the partition function using \Ref{kiri}. Then using the fact that a dual face $f^*$ is literally an edge $e\in\Delta$, the spin foam amplitude reads:
\be
Z_\Delta=
\int\prod_{e^*\in \Delta^*}d\vec{u}_{e^*}\int_{R_+}\prod_{e\in \delta} l_edl_{e} \,
\prod_{e\in \delta}\f{\sin l_{e}|\vec{u}_e|}{|\vec{u}_e|}.
\ee
Usually, a spin foam amplitude is presented as a product of weights depending solely on the $l_e$ variables after integration over the holonomies $\vec{u}_{e^*}$. Nevertheless, this first order formalism, keeping both $\vec{u}_{e^*}$ and $l_e$ variables, is necessary to describe the insertion of particles. More precisely, inserting a particle on a given edge $e$ with mass $m_e$ corresponds to the observable:
\be
{\cal O}^{(ab)}_e(m_e)
=\f{\delta(l_e)}{l_e^2}\times
\f{\delta\left(|\vec{u}_e|-m_e\right)}{4\pi m_e}.
\ee
One can obtain this expression as the abelian limit of the Ponzano-Regge spinfoam model discussed in the next section whose particle insertions have been described in \cite{PR}.
Then inserting a whole Feynman diagram corresponds to putting particles on all edges of a given graph $\Gamma$. Inserting the corresponding observable into the partition function gives:
\be
Z^{(ab)}_\Delta(\Gamma,\{m_e\})=
\int\prod_{e^*}d\vec{u}_{e^*}
\prod_{e\in\Gamma}\f{\delta\left(|\vec{u}_e|-m_e\right)}{4\pi m_e}
\prod_{e\notin\Gamma}\delta(\vec{u}_e).
\label{zgabelian}
\ee

\medskip

Now one would like to show that the previous spin foam amplitude
\Ref{zgabelian} reproduces the Feynman evaluation \Ref{ig}. The
two expressions already look very similar.  But we would like to
be more exact and identify a particular 3d triangulation $\Delta$
(or more precisely a class of triangulations) such that the
equality $Z_\Delta(\Gamma,\{m_e\})=I_\Gamma(m_e)$ holds.

Let us start by a couple of remarks on the spin foam amplitude.
First, the BF partition function is topologically invariant,
meaning that two triangulations with the same topology will yield
the same spin foam amplitude. Then, following the analysis of
\cite{gf,PR}, we know that the spin foam amplitude is in
general ill-defined and that it requires a gauge-fixing in order
to give a finite amplitude. Roughly, the origin of the divergence
is that the $\delta$ function appearing in the spin foam partition
function are redundant and, using the topological invariance, we
can generally choose a smaller set of conditions to impose that
the curvature is flat. The completely gauge-fixed amplitude for an
arbitrary triangulation $\Delta$ corresponds to the smallest
triangulation of the same topology and compatible with the
boundary data (and the graph $\Gamma$ being considered as
boundary).

The main result is that  the equality is achieved
for any 3d triangulation $\Delta$ topologically equivalent to the 3-sphere:
\be Z_{\Delta}(\Gamma,\{m_e\})=I_\Gamma(m_e). \ee A general
proof in the non-abelian case will be presented in the next
section. Let us still explain the reason why it works. The reader
can also find explicit examples in appendix.

We start with a graph $\Gamma$ embedded in $\Delta$, and we
introduce a framing i.e a surface $S\subset\Delta$ in which
$\Gamma$ is faithfully embedded.  First we consider the case with
no non-contractible cycles, $S\sim {\cal S}^2$. The triangulation
of the surface $S$ directly provides us with a triangulation of
${\cal S}_3$: ${\cal S}_3$ can be realized by taking two copies of
the 3-ball glued back together along ${\cal S}_2$. Then simply
choosing as triangulation $\Delta$ the initial triangulation for
$S$, the two expressions \Ref{ig} and \Ref{zgabelian} simply
match.

The case with cycles is a bit more subtle. Once again, we start
with the graph $\Gamma$ drawn on the surface $S$, on which we have
chosen certain sequences of dual links $\eb$ to represent the
non-contractible cycles. Looking at the Feynman evaluation, we
associate a variable $\vec{u}_C$ to each cycle. To match the spin
foam amplitude \Ref{zgabelian}, we  associate a face of the
3d triangulated manifold $\Delta$ to each cycle:
the boundary of this face being the given cycle. More precisely, the cycles of
an orientable surface $S$ can be split into a set of cycles $a_i$
and their dual cycle $b_i$. We construct the 3d spacetime by
taking two copies of $S$, one for which we add faces corresponding
to the cycles $a_i$ and the other for which we turn the cycles
$b_i$ into faces. The added faces might not be triangles at first,
but we can triangulate them in the obvious way. Gluing these two
filled-up copies of $S$ results into a 3d triangulation once again
topologically equivalent to the 3-sphere ${\cal S}_3$. The idea is
that the interior of $S$ with added faces along one set of cycles,
$a$ or $b$, is topologically a 3-ball. The point is that the
triangulation resulting from gluing the two copies of $S$ with
added faces makes the equality between \Ref{ig} and
\Ref{zgabelian} straightforward.

Here we have found one particular triangulation $\Delta_0$ for each Feynman diagram $\Gamma$
such that $\wtl{Z}_\Gamma(\Delta_0)=I_\Gamma$. Nevertheless, due to the topological
properties of the spin foam models for BF theories, it follows that the same statement is true
for any triangulation $\Delta$ topologically equivalent to $\Delta_0$ i.e which can
be constructed out of $\Delta_0$ by a sequence of Pachner moves (without modifying the graph
$\Gamma$).

Let us conclude that this equality between the spin foam amplitude
and the Feynman evaluation allows us to identify the momentum
$\vec{p}_e$ of the particle living on an edge $e$ of the
triangulation with the holonomy $\vec{u}_e=\vec{u}_{f^*}$ around
the dual face to that edge: this shows us how to encode a particle
as geometrical data.

\medskip

Let us also point out that the Feynman evaluation corresponds to
the simplest topology - the one of the 3-sphere. The spin foam
framework allows to generalize these Feynman graph amplitudes to
arbitrary topologies. It should be interesting to
understand better what effect the non trivial topology
of the ambient manifold has on Feynman graph evaluation.

In the next section, we will generalize our framework to the
non-abelian context of the Ponzano-Regge spinfoam model for 3d
quantum gravity, and analyze the Feynman evaluation which results
from inserting particles in the quantum gravity theory. This will
lead us in the following section to show that the no-gravity limit
of the Ponzano-Regge reproduces the abelian spinfoam model and the
usual classical Feynman graph evaluation and to identify the
Ponzano-Regge amplitudes as providing a perturbative expansion in
the gravity coupling which is interpreted as QFT amplitudes on a non-commutative
geometry.

\section{Particle Insertions in Ponzano-Regge Spinfoam Gravity}
\label{nonabelian}

In \cite{PR} the general form of Feynman graph amplitude for spinning particles in the
Ponzano-Regge model has been written.

In this section we focus on the case of spinless particle (a discussion of the spinning
case is included in the appendix), we first recall briefly the general construction
and then compute explicitly the
Feynman diagram amplitudes of  particles  coupled to three
dimensional Euclidean gravity in a form allowing us to take the
``no gravity'' limit i.e where the Newton constant $G_N\arr 0$.
The main result of this section (see eq. (\ref{zg1}, \ref{zg2})is a explicit computation
of the Ponzano-Regge amplitude which allow a comparison with the
usual Feynman graph amplitude computed in the previous section.
A similar result has been very recently obtained independently, in the context of
the Turaev-Viro model, by Barrett et al. \cite{Barrett2}.

We start from a triangulation $\Delta$ of our spacetime $M$ and
consider also the dual $\Delta^*$: dual vertices, edges and  faces
correspond  respectively to tetrahedra, faces and edges of
$\Delta$. We choose our Feynman graph to be embedded in the
triangulation $\Delta$ such that edges of $\Gamma$ are edges of
the triangulation. Each edge of $\Gamma$ is labelled by an angle
$\theta \in [0,\pi]$
\be
\theta= \kappa m , \qquad
\kappa= 4\pi G_N
\label{kappa}
\ee
where $G_N$ is the Newton constant, $\kappa$ is the inverse Planck mass
and $m$ is the mass of the particle.

We choose a group $G$, here $\SU(2)$, and assign group elements
$g_{e^*}$ to all dual edges $e^*$ of the triangulation.
We constrain the holonomies around dual faces $f^*\sim e$ to be flat if there is no particle
and we constraint it to be in the conjugacy class $\theta_e$
if $e$ is an edge of $\Gamma$.
More
precisely, let us note $G_{e}=G_{f^*}$ the product of the group elements
around a dual face (or plaquette) $f^*\sim e$:
$$
G_e=G_{f^*}=\prod_{e^*\in\pp f^*}g_{e^*}^{\e_{f^*}(e^*)},
$$
where $\e_{f^*}(e^*)$ records the orientation of the edge $e^*$ in
the boundary of the (dual) face $f^*$.
The amplitude is well defined once we chose a gauge fixing.
In order to do so we choose $T$ a maximal tree of $\Delta\setminus\Gamma$
and $T^*$ a maximal tree of $\Delta^*$ \cite{gf}.
Then the partition function
reads:
\be
\label{PRamp1}
 \zz_M(\Gamma_\theta)=
\int\prod_{e^*\notin T^* }dg_{e^*}\,
\prod_{e\notin T \cup \Gamma}\delta(G_{e})
\prod_{e\in \Gamma}\Delta(\theta_e) \delta_{\theta_e}(G_{e}),
 \ee
 where $dg$ is the normalized Haar measure and $\delta(g)$ the corresponding delta function on $G$
 and $\Delta(\theta)\equiv \sin(\theta)$.
The partition function contains a factor $\prod_e\Delta(\theta_e)$ which can be factor out,
this factor is important when we consider the no gravity limit. In this section in order to simplify notation we
will work with the reduced partition function
$$\wtl{\zz}_M(\Gamma, \theta) \equiv \frac{{\zz}_M(\Gamma, \theta)}{\prod_e\Delta(\theta_e)}.$$

Also, $\delta_\theta (g)$ fixes the group element $g$ to be in the
conjugacy class\footnotemark labelled by $\theta$. This fixes a
non-zero deficit angle around the edge $e$, which corresponds to
the geometrical picture of a particle in a 3d spacetime of mass
$4\pi G m=\theta$ ($e$ being the trajectory of the particle).
\footnotetext{The Cartan subgroup $H$ of $\SU(2)$ is the group of
diagonal matrices
$$
h_\theta=\left(
\begin{array}{cc}
e^{i\theta} & 0 \\
0 & e^{-i\theta}
\end{array}
\right),
$$
which correspond to rotations of angle $2\theta$ around a given axis (usually the $z$ axis).
Every group element is conjugate to such an element.
The residual action on $H=\U(1)$ is given by the Weyl group. It is $\Z_2$,
since $h_\theta$ and $h_{-\theta}$ are conjugated.} More precisely, the distribution
$\delta_\theta(g)$ is defined by:
\be
\forall f,\,
\int_G dg\, \delta_\theta(g) f(g)=
\int_{G/H}dx\, f(xh_\theta x^{-1}),
\label{delta}
\ee
where $dg$ and $dx$ are normalized invariant measures.
The amplitude \Ref{PRamp1} does not depend on the choice of the triangulation and
the gauge fixing trees, but only on the topology of the manifold $M$ and the
embedding of $\Gamma$ in it \cite{PRII}.

We can expand the $\delta$ functions in terms of characters
\bes
\delta(g)&=&\sum_j d_j \chi_j(g), \nn\\
\delta_\theta(g) &=&\sum_j \chi_j(h_\theta) \chi_j(g),\,
\mathrm{with}\, \chi_j(\theta)\equiv \chi_j(h_\theta)=\f{\sin d_j\theta}{\sin\theta},
\label{char}
\ees
$d_j =2j+1$ being the dimension of the spin $j$ representation.
We eventually  perform the integration over
$g_{e^*}$ in order to obtain a state sum model
\be\label{PRamp}
\wtl{\zz}_M(\Gamma_\theta)=\sum_{\{j_e\}} \prod_{e\notin \Gamma} d_{j_e} \prod_{e\in \Gamma}\chi_{j_e}(h_{\theta_e})
\prod_{e\in T} \delta_{j_e,0}
\prod_t
\left\{
\begin{array}{ccc}
    j_{e_{t_{1}}} &  j_{e_{t_{2}}} &  j_{e_{t_{3}}} \\
    j_{e_{t_{4}}} &  j_{e_{t_{5}}} &  j_{e_{t_{6}}}
    \end{array}
    \right\},
    \ee
where the summation is over all edges of $\Delta$ and the product of normalized 6j symbols is
over all tetrahedra $t$. For each tetrahedron, the admissible triples of edges, e.g. $(j_{e_{t_{1}}},j_{e_{t_{2}}},j_{e_{t_{3}}})$, correspond to  faces of this tetrahedra.
The factor $\delta_{j,0}$ comes from the gauge fixing, it eliminates the sum over $j_e$, $e\in T$.

Let us underline the fact that the spin foam amplitudes, like \Ref{PRamp}, are purely algebraic constructions built with dimensionless quantities. The gravitational coupling constant $G_N$ does not directly enter into the partition function. It appears as a unit used to turn the dimensionless quantities (like the angle $\theta$) into physical properties of the matter and particles (like the mass $m$).

\medskip

Starting from the definition (\ref{PRamp}) we can  explicitly compute
these amplitudes.
First, in order to make contact with usual Feynman graph amplitudes
we have to restrict the topology of the spacetime to be trivial, so
$M= {\cal S}^3$.
Also, to warm up, we first consider the case where $\Gamma$ is a spherical graph
which can be embedded in ${\cal S}^2$.
Lets start with the example of $\Gamma$ being a tetrahedral graph
embedded in ${\cal S}^2\subset {\cal S}^3$.
We denote by $I=1,..,4$ the vertices of this graph and
$e=(IJ)$ the edges of this graph.
Since the amplitude does not depend on the triangulation
we are free to chose it. The simplest triangulation of
${\cal S}^3$ we can chose in which the graph can be embedded
consist of two tetrahedra.
One of the tetrahedra gives a triangulation of the interior of
${\cal S}^2$ the one other gives a triangulation of the
exterior of ${\cal S}^2$ the three sphere is obtained by gluing the two balls together.
With this graph and triangulation, no gauge fixing is needed,
 and the corresponding amplitude reads
\be\label{PRamptet}
\wtl{\zz}_{\S^3}(\Gamma_\theta)=\sum_{\{j_{IJ}\}} \prod_{I<J}\chi_{j_{IJ}}({\theta_{IJ}})
\left\{
\begin{array}{ccc}
    j_{34} &  j_{24} &  j_{23} \\
    j_{12} &  j_{13} &  j_{14}
    \end{array}
    \right\}^2.
\ee
The $6j$ square can be written as a group integral
\be\label{groupint}
\left\{
\begin{array}{ccc}
    j_{34} &  j_{24} &  j_{23} \\
    j_{12} &  j_{13} &  j_{14}
    \end{array}
    \right\}^2
    =\int \prod_I dg_I \prod_{I<J} \chi_{\bar{\jmath}_{IJ}}(g_Ig_J^{-1}),
    \ee
where we have introduce the notation $\bar{\jmath}_{IJ}\equiv j_{KL}$, with $I,J,K,L$ all distinct.
$\bar{\jmath}_{IJ}$ label the edges of the graph dual to $\Gamma$ in $\S^2$.
We also denote $ \bar{\theta}_{IJ} = \theta_{KL}$.
Using the previous evaluation and character expansion (\ref{char})
we can perform the summation over spins and obtain
\be
\wtl{\zz}_{\S^3}(\Gamma_\theta)= \int\prod_I dg_I \prod_{I<J} \delta_{\bt_{IJ}}(g_Jg_I^{-1}).
\ee
This can be explicitly evaluated \cite{asymp}
\be
{\zz}_{\S^3}(\Gamma_\theta)=\wtl{\zz}_{\S^3}(\Gamma_\theta) \prod_{I<J}\sin(\theta_{IJ})=
\f{\pi^2}{2^5}\f{1}{\sqrt{\mathrm{det}(\cos\theta_{IJ})}}.
\ee

Lets  now consider a general spherical graph $\Gamma \subset \S^2$.
In general $\Gamma $ is not a triangulation of $\S^2$ but we can add edges to it
 in order to have a regular triangulation of $\S^2$,
we denote it ${\Gamma}_{\Delta}$.
Once this is done we write as before $\S^3$ as the gluing of two 3-balls, one being the interior of
$\S^2$ the other the exterior.
We can extend the triangulation ${\Gamma}_{\Delta}$ of $\S^2$ to a triangulation of the interior ball
by adding interior edges and interior vertices and gluing corresponding tetrahedra\footnote{For instance, let's choose an order of the vertices $V=0,1,\cdots,n$ such that vertices $I$ and $I+1$ belong to
neighboring triangles (sharing an edge). We choose $0$ as our reference vertex,
then we consider the vertex $1$. If $01$ is already an edge of our boundary triangulation we
do nothing and consider the vertex $2$. If $01$ is not an edge of our triangulation we
add it in the interior of the ball and consider the tetrahedron which consists of this new edge and
of the two neighbor triangles containing $0$ and $1$. And then we continue the procedure.
This will provide a triangulation of the 3-ball which extends the triangulation of $S^2$ and possesses
no interior vertex.}.
We can choose the triangulation of the exterior ball to be the same as the interior ball with reversed orientation.

Starting from the amplitude (\ref{PRamp}), we can perform the summation over all edges of the triangulation that do not belong
to $\S^2$, we obtain
\be\label{PRampsphere}
\wtl{\zz}_{\S^3}(\Gamma_\theta)=
\sum_{\{j_e\}}\prod_{e\notin \Gamma} d_{j_{e}} \prod_{e\in \Gamma}\chi_{j_e}(h_{\theta_e})
|\bra B^3|\Gamma_\Delta,j_e\ket|^2
\ee
where the summation is only over boundary edges and
  $\bra B^3|\Gamma,j_e \ket$ is the Ponzano-Regge amplitude associated with the 3-ball,
with $(\Gamma,j_e)$ on its boundary\footnote{
We can extend the definition of Ponzano-Regge amplitude to the case of manifold
with boundary $\partial M =\Sigma$ with a colored graph drawn on the boundary triangulation to be
\be
\bra M|\Gamma,j_e)= \sum_{\{j_e\}}\prod_ed_{j_e} \prod_{e\in T}\delta_{j_e,0}\prod_t
\left\{
\begin{array}{ccc}
    j_{e_{t_{1}}} &  j_{e_{t_{2}}} &  j_{e_{t_{3}}} \\
    j_{e_{t_{4}}} &  j_{e_{t_{5}}} &  j_{e_{t_{6}}}
    \end{array}
    \right\}.
    \ee
    where the summation is only over internal edges.}.
    This amplitude can be understood as the physical scalar product between a spin network state
    and a `Hartle-Hawking' state associated with a manifold with boundary.
This amplitude can be easily computed and is given by the evaluation of the corresponding spin network.
More precisely, let's
denote $\overline{\Gamma}$ the graph dual to $\Gamma$
in $\S^2$ and $\overline{\Gamma}_\Delta$ the graph dual in $\S^2$ to ${\Gamma}_\Delta$.
The edges of $\overline{\Gamma}_\Delta$ are denoted $\bar{e}$ and are in one to one correspondence with edges
of ${\Gamma}_\Delta$.
Given a coloring $j_{\bar{e}}=j_{e}$ of the edges of $\overline{\Gamma}_\Delta$
we can consider the spin network functional
\be
 \Phi_{(\overline{\Gamma}_\Delta,j_{\bar{e}})}(g_{\bar{e}})
 \ee
 which is a gauge invariant function on $G^{|E|}$ with the gauge group acting at vertices.
 Then
\be
\bra B^3|\Gamma_\Delta,j_e\ket = \Phi_{(\overline{\Gamma}_{\Delta},j_{\bar{e}})}(1).
\ee
The modulus square of this amplitude can be expressed as an
integral
\be
|\Phi_{(\overline{\Gamma}_{\Delta},j_{\bar{e}})}(1)|^2=\int \prod_{\bar{v}}dg_{\bar{v}} \prod_{\bar{e}}
\chi_{j_{\bar{e}}}(g_{t(\bar{e})}g_{s(\bar{e})}^{-1}),
\ee
where $s(\bar{e}),t(\bar{e})$ denote the starting and terminal vertices of the oriented edge
$\bar{e}$.
We can now perform the summation over $j_{\bar{e}}$ in (\ref{PRampsphere})
in order to get
\be
\wtl{\zz}_{\S^3}(\Gamma_\theta)= \int \prod_{\bar{v}} dg_{\bar{v}}
\prod_{\bar{e}\in \overline{\Gamma}} \delta_{\theta_{\bar{e}}}(g_{t(\bar{e})}g_{s(\bar{e})}^{-1})
\prod_{\bar{e}\notin \overline{\Gamma}} \delta(g_{t(\bar{e})}g_{s(\bar{e})}^{-1}).
\ee
We can easily integrate out all the delta functions associated
with edges of $\overline{\Gamma}_\Delta$ not in $\overline{\Gamma}$
in order to finally obtain
\be
\wtl{\zz}_{\S^3}(\Gamma_\theta)= \int \prod_{\bar{v}\in \Gamma} dg_{\bar{v}}
\prod_{\bar{e}\in \Gamma}\delta_{\theta_{\bar{e}}}
(g_{t(\bar{e})}g_{s(\bar{e})}^{-1}),
\label{zgsphere}
\ee
which is the desired result.

We now consider the general case of a graph
which is not necessarily spherical but which can be embedded in a
Riemann surface of genus $g$, $\Gamma \subset \Sigma_g$.
As before we can add edges to $\Gamma$ in order
to have a regular triangulation of $\Sigma_g$,
denoted ${\Gamma}_{\Delta}$.
Once this is done we write  $\S^3$ as the gluing of two
handlebodies of genus $g$: $\S^3= H_g \sharp_{\Sigma_g} H_g^*$.
The meridians of the interior handlebody
draw a set of $a_i(i=1,\cdots,g)$ cycles on $\Sigma_g$,
the meridians of the exterior handlebody draw a set of
 $b_i(i=1,\cdots,g)$  cycles on $\Sigma_g$.
 The $a$ cycles intersect transversally the $b$ cycles and together they form
 a base of $H^1(\Sigma_g)$.
We can extend the triangulation ${\Gamma}_{\Delta}$ of $\Sigma_g$ to a triangulation of the interior
handlebody $H_g$.
In order to do so we first have to choose a representative of each meridian
as a cycle of edges in ${\Gamma}_{\Delta}$.
Each edge $e\in {\Gamma}_{\Delta}$  belongs to one of the cycle $a_i$
or to none.
We introduce an index $\imath(e)$ to keep track of this,
where $\imath(e)\in 0,1,\cdots, g $ is equal to $0$ if it doesn't belong to
any cycle and it is equal to $i$ if it belongs to the cycle $a_i$.
Each meridian $a_i$ is the boundary of a meridian disk $D_i$ cutting the handlebody.
We choose a triangulation $\Delta_i$ of $D_i$ which matches the triangulation
of $a_i$ on its boundary.
Once this is done we cut $H_g$ along the disks $D_i$
and consider $H_g -\cup_i D_i$. This is a three ball, its boundary
contains 2 copies of each meridian disk $D_i$ which have to be identified to reconstruct
$H_g$.
Moreover, the triangulation $\Gamma_\Delta$ and the triangulation of the meridian disk
induces a triangulation $\Gamma_\Delta \cup \Delta_i \cup \Delta_i^*$ of the boundary $\S^2$ of this ball.
We can, as previously, extend this 2-dimensional triangulation to a three dimensional triangulation
 of the ball.
Gluing back the ball along $D_i$ we obtain a triangulation of $H_g$.
For the exterior handlebody $H_g^*$ we do the same except that we have to exchange
the cycles $a_i$ with the cycles $b_i$.
In this case we denote $\tilde{\imath}(e)$ the index specifying which $b$ cycle
$e$ belongs to.

Starting from the amplitude (\ref{PRamp}) we can
perform the summation over all edges of the triangulation that do not belong
to $\Sigma_g $, we obtain
\be\label{PRampsphere2}
\wtl{\zz}_{\S^3}(\Gamma_\theta)=
\sum_{\{j_e\}}\prod_{e\notin \Gamma} d_{j_{e}} \prod_{e\in \Gamma}\chi_{j_e}(h_{\theta_e})
\bra \Delta_\Gamma,j_e|H_g^* \ket  \bra H_g|\Delta_\Gamma,j_e\ket
\ee
where the summation is only over boundary edges and
$\bra H_g|\Gamma,j_e\ket$ is the physical scalar product between the spin network state
$|\Gamma,j\ket$ and the `Hartle-Hawking' state $ |H_g\ket$.
This scalar product can be computed, it is given by
\be\label{Hgamp}
\bra H_g|\Gamma_\Delta,j_e\ket =\int
\prod_{i=1}^g da_i  \, \Phi_{(\overline{\Gamma}_{\Delta},j_{\bar{e}})}(a_{\imath(\bar{e})}^{\epsilon(\bar{e})})
\ee
where, $\Phi_{(\overline{\Gamma}_{\Delta},j_{\bar{e}})}(g_{\bar{e}})$ is the spin network functional,
$\imath(\bar{e})$ label which cycle $e$ belongs to, it is $0$ if it belongs to none,
and $a_0 \equiv 1$.
$\epsilon(\bar{e})=+1$ if $e $ as the same orientation than $a_i$ and $\epsilon(\bar{e})=-1$
otherwise.
The proof of this evaluation goes as follows:
the idea is to express, in terms of amplitudes, the fact that $H_g$ is the gluing of
a ball along the meridians disks.
The edges of the triangulation $\Gamma_{\Delta}$ of $H_g$ are colored by spins $j_e$,
the triangulation $\Delta_i$ of $D_i$ carries additional edges $\tilde{e}_i$ colored by $j_{\tilde{e}_i}$.
The triangulation of the ball
$\Gamma_\Delta \cup \Delta_i \cup \Delta_i^*$ is then colored by spins $j_e,j_{\tilde{e}_i}$ with the
additional condition that the spin coloring $\Delta_i$ and $\Delta_i^*$ are the same.
The handlebody amplitude is then obtained by summing the ball amplitude over all spins $j_{\tilde{e}_i}$
\be\label{boundampg}
\bra H_g|\Gamma_\Delta, j_e\ket
= \sum_{j_{\tilde{e}_i}}d_{j_{\tilde{e}_i}}
\bra B^3| \Gamma_\Delta \cup \Delta_i \cup \Delta_i^*,j_{e},j_{\tilde{e}_i}\ket
\ee
We know that
$\bra \S^3| \Gamma_\Delta \cup \Delta_i \cup \Delta_i^*,j_{e},j_{\tilde{e}_i}\ket$
is just given by the evaluation of a spin network based on a graph dual to
$\Gamma_\Delta \cup \Delta_i \cup \Delta_i^*$.
Moreover, this spin network is such that each spin $j_{\tilde{e}_i}$ appears
twice. We can therefore express the evaluation as an integral
over group elements associated to the disk's vertex $\tilde{v}$ which are in the interior of the $D_i$'s.
\be
\bra \S^3| \Gamma_\Delta \cup \Delta_i \cup \Delta_i^*,j_{e},j_{\tilde{e}_i}\ket
=\int \prod_{\tilde{v}}dg_{\tilde{v}}
\Phi_{(\overline{\Gamma}_{\Delta},j_{\bar{e}})}(g_{\tilde{v}(e)})
\prod_{\tilde{e}_i} \chi_{j_{\tilde{e}_i}}(g_{t_{\tilde{e}_i}}g_{s_{\tilde{e}_i}}^{-1}),
\ee
where $\tilde{v}(e)$ is the disk's vertex which belong to $e$. If $e$ doesn't intersect any
meridian  disk it is understood that $g_{\tilde{v}(e)}=1$.
Now the summation over the spins in (\ref{boundampg})
produces a $\delta$ function for every disk edge.
One $\delta$ function per disk is eliminated by the gauge fixing.
 We can then integrate out all
the group element $g_{\tilde{v}}$ except one per disk
which we call $a_i$.
This gives us the expected formula (\ref{Hgamp}).
If we insert this evaluation into (\ref{PRampsphere2}) we get
\be\label{stringlike}
 \wtl{\zz}_{\S^3}(\Gamma_\theta)=\int \prod_{i=1}^g da_i
db_i\, \sum_{\{j_e\}}\prod_{e\notin \Gamma} d_{j_{e}} \prod_{e\in
\Gamma}\chi_{j_e}(h_{\theta_e})
 \Phi_{(\overline{\Gamma}_{\Delta},j_{\bar{e}})}(a_{\imath(\bar{e})}^{\epsilon(e)})
  \overline{\Phi}_{(\overline{\Gamma}_{\Delta},j_{\bar{e}})}(b_{\tilde{\imath}(\bar{e})}^{\epsilon(e)}).
\ee
This expression is somehow reminiscent of a string theory amplitude\footnote{ In the case of string theory the
  genus $g$ amplitude is written as
  \be
  Z= \int_{{\cal M}_g}  dm d\bar{m} \sum_I \Psi_I(m)
  \overline{\Psi}{^I}(\bar{m}),
  \ee
  where the integral is over the moduli space of Riemann surface
  and $m$ is the holomorphic moduli and $I$ labels the space of
  Holomorphic Virasoro conformal blocks.
  The expression (\ref{stringlike}) have the same general form
  if we exchange the moduli $m, \bar{m}$ by $a_i,b_i$, the
  label $I$ by spin label and the Conformal block $\Psi$ by the
  spin network functional $\Phi$.
  It may be only an accidental analogy.}.
We can now express the product  $\Phi \overline{\Phi}$ as an
integral over a product of characters for each edge of $\Gamma_\Delta$ and then perform the
summation over the $j_{e}$ in order to obtain
\be
\wtl{\zz}_{\S^3}(\Gamma_\theta)=
\int\prod_{f\subset\Sigma_g}dg_f
\int \prod_{i=1}^g da_i db_i\,
\prod_{e\in \Gamma}
\delta_{\theta_e}(g_{t_{e}} a_{\imath(e)}^{\epsilon(e)}g_{s_{e}}^{-1} b_{\tilde{\imath}(e)}^{\epsilon(e)})
 \prod_{e\notin \Gamma}
\delta(g_{t_{e}} a_{\imath(e)}^{\epsilon(e)}g_{s_{e}}^{-1}
b_{\tilde{\imath}(e)}^{\epsilon(e)}), \ee where we have switched
all the notations back to the triangulation so that $t_e$ and
$s_e$ are the two faces adjacent to the edge $e$ (they
equivalently are the dual vertices $\vb$ ending the dual edge
$\eb$). If one chooses  the cycles $a,b$ to lie entirely in
$\Gamma$, we can integrate out the delta functions associated with
edges not in $\Gamma$ and obtain the result we are looking for:
\be \wtl{\zz}_{\S^3}(\Gamma_\theta)= \int\prod_{f\subset\Sigma_g}dg_f
\int \prod_{i=1}^g da_i db_i\, \prod_{e\in \Gamma}
\delta_{\theta_e}(g_{t_{e}}
a_{\imath(e)}^{\epsilon(e)}g_{s_{e}}^{-1}
b_{\tilde{\imath}(e)}^{\epsilon(e)}). \label{zg1} \ee This is the
generalization of the Feynman evaluation \Ref{ig} to the
non-abelian case when inserting particles in 3d quantum gravity.
Further assuming that each edge $e$ belongs to a unique cycle of
$\Sigma_g$, one can simplify the formula :
\be
\wtl{\zz}_{\S^3}(\Gamma_\theta)= \int\prod_{f\subset\Sigma_g}dg_f \int
\prod_{C\textrm{ cycles of } S} dU_C\, \prod_{e\in \Gamma}
\delta_{\theta_e}\left(g_{t_{e}} U_{C(e)}^{\e_C(e)} g_{s_{e}}^{-1}
\right). \label{zg2}
\ee

\section{The no-gravity limit and a perturbative expansion in $G$}

\subsection{The Quantum Field Theory limit of Spin Foams at $G\arr 0$}

Now we would like to study the "no-gravity" limit of the particle
insertion amplitudes of the Ponzano-Regge model. That is the limit
when we take the Newton constant to zero i.e. $G_N\arr 0$. In three
spacetime dimensions, the Planck length reads $l_P\sim\hbar G_N$
while the Planck mass is $m_P\sim 1/G_N$. The usual classical limit
is taking $\hbar\arr 0$ while keeping $G_N\ne 0$. As $m_P\ne 0$ in
this limit, we get an effective deformed Poincar\'e algebra
\cite{dsr} and recover the framework of Doubly Special Relativity.
Here, we consider the alternative limit $G_N\arr 0$ while $\hbar$ is
kept fixed, in this limit, we expect to recover from a quantum
gravity theory the usual quantum field theory framework: we call
it the {\it QFT limit}. More exactly, we want to recover the usual
Feynman diagram evaluations of quantum field theory on a flat
background as described in the section \ref{abelian}.

More precisely, the limit is defined as $\kappa\arr 0$ where
$\kappa$ is the inverse Planck mass defined in \Ref{kappa} as the
ratio between the deficit angle and the corresponding mass of a
particle. To understand how the different quantities gets
renormalized in the limit we parametrize  $g\in\SU(2)$ in terms of
a Lie algebra element $\vec{u}\in \mathbb{R}^3$. In the
fundamental representation we have: \be g\equiv e^{i\kappa
u^i\sigma_i}=\cos(m\kappa)+i\sin(m\kappa)\vec{n}.\vec{\sigma}, \ee
where $m=|\vec{u}|$, $\vec{n}$ is the direction of the rotation
and $\vec{\sigma}$ the Pauli matrices. in the limit, the angle $\theta=m\kappa$
goes to 0 so that we are in fact considering perturbations around
the identity in $\SU(2)$ or equivalently the limit in which the
group $\SU(2)$ goes flat and becomes the abelian group defined by
its Lie algebra $\su(2)$. So it is natural to also call the QFT
limit the {\it abelian limit} for spin foams.

We can expand group elements
and their products perturbatively in the parameter $\kappa$:
$$
g_i\sim\left(1-\kappa^2\frac{|\vec{u}_i|^{2}}{2}\right)+i\kappa \vec{u}_i\cdot\vec{\sigma}+\dots,
$$
\be
g_1g_2\sim 1+i\kappa\left(\vec{u}_1+\vec{u}_2\right)\cdot\vec{\sigma}
+\f{\kappa^2}{2}\left(
\vec{u}_1^2+\vec{u}_2^2+2\vec{u}_1\cdot \vec{u}_2
+2i (\vec{u}_1\wedge \vec{u}_2)\cdot\vec{\sigma}
\right)+\dots,
\ee
so that the group multiplication is linear at the first order in $\kappa$.
In the $\kappa \arr 0$ limit, the discrete representations label  $j$ becomes a continuous length parameter
$l$ parameterizing the representations of $\su(2)\sim \R^3$ as a group and $m$ is the renormalized mass.
The QFT limit is defined as by: \be
d_j=\f{l}{\kappa}, \qquad \theta=m\kappa,
\qquad \kappa\arr 0,\qquad l,m\,\, \mathrm{fixed} \ee
In this limit the integral over the group becomes integral over the Lie algebra
\be
\int_G dg \sim_{\kappa\arr 0} \kappa^3\int_{\R^3} \frac{d^3\vec{u}}{2\pi^2}
\ee
where the normalization factor insures that
\be
\int \frac{d^3\vec{u}}{2\pi^2} f(\vec{u}) = \frac{2}{\pi}\int_0^\infty dm\, m^2 \int_{{\cal S}^2}d^2\vec{n} f(m\vec{n}),
\ee
with $d\vec{n}$  the normalized measure on the sphere\footnote{The
 normalization comes from the Weyl integration formula
 \be \int dg f(g)= \frac{2}{\pi} \int_0^\pi d\theta \sin^2(\theta) \int_{G/H}dx f(xgx^{-1})
 \ee with $dg,dx$ being normalized invariant measures}.
Summation over $j$ becomes an integral
\be
\sum_{j}  \sim_{\kappa\arr 0} \frac{1}{\kappa}\int_0^\infty dl,
\ee
and we recover the usual classical Hadamard propagator \Ref{usual} as the
abelian limit of the $\SU(2)$ character.
\be
\chi_j(g)=\f{\sin d_j\theta}{\sin\theta}\sim_{\kappa\arr 0}
\frac{1}{\kappa} \f{\sin l |\vec{u}|}{|\vec{u}|}.
\ee
The Ponzano-Regge partition function (\ref{PRamp1}) is given by
\be
 \zz_M(\Gamma_\theta)=
\int\prod_{e^*\notin T^* }dg_{e^*}\,
\prod_{e\notin T \cup \Gamma}\delta(G_{e})
\prod_{e\in \Gamma}\delta_{\theta_e}(G_{e})\Delta(\theta_e).
 \ee
Let's first consider the case of a closed manifold without
particles and expand the delta functions in terms of characters, this gives
\be
\zz_M=\sum_{j_e}
\prod_{e }{d_{j_e}}\int\prod_{{e^*}}dg_{{e^*}}\,
\prod_{e}\chi_{j_{e}}(G_e).
\label{PRvol1}
\ee
Now taking the limit $\kappa\arr 0$ is straightforward.
One uses the fact that the product $g_e=g_{f^*}=\prod_{e^*}g_{e^*}$ is an abelian product
at the first order in $\kappa$ then takes the limit of the characters. This leads to:
\be
Z= \lim_{\kappa\arr0}\zz
=\int\prod_{e^*\notin T^*}\frac{d^3\vec{u}_{e^*}}{2\pi^2}\int_{R_+}l_edl_{e}
\prod_{e\notin T}\f{\sin l_{e}|\vec{u}_e|}{|\vec{u}_e|}
=\int_{\R^3}\prod_{e^*}{d\vec{u}_{e^*}}
\prod_{e}\delta(\vec{u}_e),
\ee
where the vectors $\vec{u}_{e^*}$ are the Lie algebra counterpart of  the group elements $g_{e^*}$:
$$
\vec{u}_{e^*}\equiv m_{e^*} \vec{n}_{e^*},
\qquad
\vec{u}_{e}\equiv \vec{u}_{f^*}\equiv \sum_{e^*\in\pp f^*} \vec{u}_{e^*}.
$$
We see that we naturally recover the abelian model studied previously.
In the limit we have to keep track of the factor $\kappa$.
For each edge $e$ not in $T$, there is a factor $1/\kappa^3$ (one coming from $\sum_j$, one from $d_j$ and one
from $\chi_j$) and for each dual edge $e^*$ not in $T^*$ there is a factor $\kappa^3$.
Overall this means that $\kappa$ comes with the power $3\chi(M)$, where
$\chi(M)=\sigma_0-\sigma_1+\sigma_2-\sigma_3$ is the Euler characteristic of $\Delta$
and  $\sigma_{i}$ are the number of i-cells
of the triangulation.
For a closed three manifold $\chi(M)=0$ so no factor $\kappa$ appears in the limit of the partition
function.

A particle insertion of  mass $\theta_e$ on an edge $e$, or dual face ${f^*}$,
corresponds to inserting the following observable \cite{PR} in the partition function (\ref{PRvol1}):
\be\label{parop}
{\cal O}_{e}(\theta_e)=
\delta_{j_e0}\times \delta_{\theta_e}(g_e)\times \Delta(\theta)
\ee
where  $\Delta(\theta)=\sin\theta$ and the distribution $\delta_\theta(g)$ is defined in (\ref{delta}).
Inserting this operator in $\zz$ relaxes the constraint $\delta(g)$ by replacing it by the weaker
constraint $\delta_\theta(g)$, which gives back the formula \Ref{PRamp1}.

For a particle insertion on the edge $e$ with mass $m_e$, we first observe that
 the limit of $\delta_\theta(g)$ is $\delta_m(\vec{u})$, which is a distribution satisfying
the condition (obtained as the limit of \Ref{delta}):
\be
\int_{\R^3}\frac{d^3\vec{u}}{2\pi^2}\,\delta_m(|\vec{u}|)f(\vec{u})=\frac{1}{\kappa^3}
 \int_{{\cal S}^2} d^2\vec{n}\,f(m\vec{n})
\ee so $\delta_m(\vec{u})= \frac{\pi}{2\kappa^3 m^2}
\delta(|\vec{u}|-m)$. Note that we have the analog of identity
(\ref{char})
\be \int_0^\infty dl
\frac{\sin{lm}}{m}\frac{\sin{l}|\vec{u}|}{|\vec{u}|} =
\frac
\pi{2m^2} \delta(|\vec{u}|-m).
\ee
The term $\delta_{j0}$ in
(\ref{parop}) kills the summation $\sum_j d_j \chi_j(G)$ which
becomes $1/\kappa^3 \int dl\, l \frac{\sin{l|\vec{u}|}}{|\vec{u}|}$
in the abelian limit. So the abelian limit of $\delta_{j0}$ is $\kappa^3\frac{\delta(l)}{l^2}$.
Eventually, the abelian limit of $\Delta(\theta) $ is $\kappa m$.
 This shows that  the particle insertion in
the $\kappa\arr 0$ corresponds to the following observable of the abelian
case:
\be {\cal O}_e(m_e) =\kappa \f{\delta(l_e)}{l_e^2}\times
\f{\pi}{2m}{\delta\left(|u_e|-m_e\right)}.
\ee
Finally, putting everything together, the QFT limit of the Ponzano-Regge amplitude
with a particle graph $\{\Gamma,m_e\}$ is simply equal to the
amplitude for particle insertions in the abelian partition
function and reads:
\bes  Z_{\Delta}^{(ab)}(\Gamma,\{m_e\})
&=&\lim_{\kappa\arr 0} \kappa^{|e_\Gamma|}\zz_\Gamma(\theta_e)\nn\\
&=&\int\prod_{e^*}d\vec{u}_{e^*}
\prod_{e\in\Gamma}\f{\delta\left(|u_e|-m_e\right)}{4\pi m_e
}
\prod_{e\notin\Gamma}\delta(u_e).
\label{zg}
\ees
where $|e_\Gamma|$ is the number of edges in $\Gamma$.
\medskip

This shows that the QFT limit $\kappa\arr0$ of the
Ponzano-Regge amplitude  reproduces the usual classical Feynman
graph evaluation (with Hadamard propagators), which was
previously shown to be equal to the amplitudes of the abelian
$\R^3$ spin foam model of section \ref{abelian}. The next step is
to expand the non-abelian spin foam amplitudes beyond the first
order in $\kappa$ in order to understand the structure of the
Feynman graph evaluation in the non-abelian context or, in other
words, extract the quantum gravity correction to the classical QFT
Feynman graph evaluations.

\subsection{Expansion in $G$: Non-Commutative Geometry and Star Product}

\sss{Feynman evaluation of spherical graphs}

We start with the Ponzano-Regge spin foam amplitude  for a
particle graph written in section \ref{nonabelian}. We first deal with the case
of a spherical graph $\Gamma$.
Applying the (topology) duality transform described in section \ref{abelian},
we can write this amplitude exactly as a Feynman graph evaluation
similar to \Ref{evaluation}.

More precisely, if $\Gamma$ is  a spherical graph $\Sigma_g\sim{\cal S}^2$, there is no
cycle variables $a_i,b_i$ to integrate over and  the spin foam
amplitude \Ref{zg1} simplifies down to \Ref{zgsphere}.
Then let's denote  $G_e=g_{t(e)}g_{s(e)}^{-1}$ the group element associated
to each edge $e\in\Gamma$, where $t(e),s(e)$ label the two triangles bounding $e$.
It is clear that the product of such
$G_e$ variables  around any given vertex $v\in\Gamma$  is constrained to be the
identity. And in the end, we obtain the following Feynman graph
evaluation: \be \ii(\Gamma,\theta) \equiv
\zz_{\S^3}(\Gamma_\theta) =\int_{\SU(2)^E} \prod_{e \in \Gamma} dG_e\,
\Delta(G_e)\delta_{\theta_e}(G_e)\, \prod_v \delta(G_v), \ee with
$$
G_v=\overrightarrow{\prod_{e\supset v}}G_e^{\e_v(e)},
$$
where $\e_v(e)=\pm$ whether $v$ is the target or source vertex of the edge $e$.
Now one can re-express the $\delta$ function on the group $SO(3)$ as\footnotemark:
\be
\delta(G)=\f{1}{8\pi\kappa^3}\int_{\su(2)}d^3X\,e^{\f{i}{2\kappa}\,Tr(Xg)}.
\ee
\footnotetext{This formula gives us the delta function on $SO(3)=SU(2)/\Z_2$,
where $\Z_2$ is the identification $g=-g$.
In the following we restrict to the case where all group elements are
$SO(3)$ group elements for simplicity.
As pointed out in \cite{PR} if we want to construct
the delta function on $SU(2)$ the correct formula is:
$$
\delta(G)=\f{1}{8\pi}\int_{\su(2)}d^3X\,e^{\f{i}{2}\,Tr(Xg)}(1+\e(g)),
$$
where $\e(g)$ is the sign of $\cos\theta$ with
$g=\cos\theta+i\sin\theta=\hat{n}\cdot\vec{\sigma}$.}
Therefore the non-abelian Feynman evaluation reads:
\be
\ii(\Gamma,\theta)=
\int\prod_v \f{d^3X_v}{8\pi\kappa^3}\,
\int\prod_e dG_e\,
\Delta(G_e)\delta_{\theta_e}(G_e)\,
\prod_v e^{\f{i}{2\kappa}\,Tr(X_vG_v)}.
\label{PRfeyn}
\ee
If the product $G_v=\prod_{e\in\pp v}G_e$ was abelian, we
could expand the exponent and reorganize the product over
vertices into a product over edges just like the usual Feynman
graph evaluation \Ref{abevaluation}. Nevertheless, we know that
the product is actually abelian at the first order in $\kappa$,
so that we can do a perturbative expansion in $\kappa$ and compute
deviations from the classical Feynman diagram expressions.

Denoting $2i\kappa \vec{P}=Tr(g\vec{\sigma})$ the projection of the group
element $g$ on the Lie algebra, the product of $n$ group elements $g_1,..,g_n$
admits a simple expansion in terms of $\kappa$:
\be
\prod_i g_i\,=\,
1 +i\kappa \sum_i \vec{P}_i \cdot\vec{\sigma}
 -\f{\kappa^{2}}{2}\left|\sum_i \vec{P_i}\right|^2
-i\kappa^{2} \left(\sum_{i<j}\vec{P}_i\wedge\vec{P}_j \right)\cdot \vec{\sigma}
+\dots
\ee
Then we can expand the exponent in the graph amplitude:
\be
\f{i}{2\kappa}Tr(X_vG_v)=-i\vec{X}_v.\left(
\sum_{e\in \pp v} \e_v(e)\vec{P}_e
+\kappa\e_v(e)\e_v(e')\sum_{e<e'\in\pp v}\vec{P}_{e}\wedge\vec{P}_{e'}+\dots
\right),
\ee
where $X_v=i\vec{X_v}.\vec{\sigma}$. So the full Feynman evaluation reads:
\bes
\ii(\Gamma,m)=
\int\prod_v \f{d^3X_v}{8\pi\kappa^3}\,
\int\prod_e dG_e\,
\Delta(G_e)\delta_{\kappa m_e}(G_e)\,
\prod_v e^{-i\vec{X}_v.\left(
\sum_e \e_v(e)\vec{P}_e  +\kappa\e_v(e)\e_v(e')\sum_{e,e'}\vec{P}_{e}\wedge\vec{P}_{e'}+\dots\right)} \nn\\
=
\int\prod_v \f{d^3X_v}{8\pi\kappa^3}\,
\int\prod_e dG_e\,
\Delta(G_e)\delta_{\kappa m_e}(G_e)\,
\prod_v \left(1-i\kappa\e_v(e)\e_v(e')\vec{X}_v.\sum_{e,e'}\vec{P}_{e}\wedge\vec{P}_{e'}+\dots\right)
e^{-i \vec{X}_v.\sum_e \e_v(e)\vec{P}_e}
\ees
We also need to expand the measure:
$$
\int\prod_e dG_e\,\Delta(G_e)\delta_{\kappa m_e}(G_e)\, \varphi(G_e)\equiv
\int\prod_e \f{\kappa d^3\vec{P}_e}{2\pi}\,
\delta\left(|\vec{P}_e|^2-\left(\f{\sin m_e\kappa}{\kappa}\right)^2\right)\, \varphi(\vec{P}_e),
$$
in order to  express everything in terms of the moment vectors $\vec{P}_e$ and get the full perturbative expansion in $\kappa$. This leads to a simple renormalisation of the mass $m\rightarrow \f{\sin\kappa m}{\kappa}$, so that
the measure doesn't contribute at first order in $\kappa$.
If one introduces a source $\vec{J}_e$ for the variables $\vec{P}_e$, one can generate
the polynomial terms in $\vec{P}$ through some derivative with respect to the source $J$:
\be
\ii(\Gamma,m)=
\left.\int\prod_v \f{d^3X_v}{8\pi\kappa^3}\,
\left(
1+i\kappa\vec{X}_v.\sum_{e,e'\in\pp v}\e_v(e)\e_v(e')
\f{\delta}{\delta\vec{J}_e}\wedge\f{\delta}{\delta\vec{J}_{e'}}+\dots
\right).
I(\Gamma,m_e,\vec{X}_v,\vec{J}_e) \right|_{J= 0},
\ee
where $I(\Gamma,m,\vec{X}_v,\vec{J})$ is a normal (abelian) Feynman graph evaluation:
\be
I(\Gamma,m,\vec{X}_v,\vec{J})=
\int\prod_e
\f{\kappa d^3\vec{P}_e}{2\pi}\,
\delta\left(|\vec{P}_e|^2-\f{\sin^2 m_e\kappa}{\kappa^2}\right)\, \varphi(\vec{P}_e)
\prod_e e^{-i\kappa \vec{P}_e.(\vec{X}_{t(e)}-\vec{X}_{s(e)})-i\vec{J}_e.\vec{P}_e}.
\ee
Therefore we see that the Feynman graph
evaluation for particles in 3d  quantum gravity
can be expressed as a perturbative expansion in
$\kappa$  with operators acting on a
classical Feynman graph evaluation.

At first order the operator to insert is a sum of operator
acting at the vertices of the Feynman diagram which modify how
the particles interact.
At higher orders the measure will also contribute and modify the
propagator of the theory.

One can compute the higher order correction and this perturbative expansion looks quite cumbersome at first sight. However we are now going to see how the full
perturbative expansion can be simplified if one
re-express the previous formulas in terms of a
{\it star product}, which renders explicit the non-commutative structure of the
theory and scattering amplitudes.

\sss{Star Product and Non-Commutative Field Theory}

Starting from the non-abelian Feynman evaluation \Ref{PRfeyn}, one
would like to split the exponentials $\exp(\f{i}{2}Tr(X_v\prod_e G_e))$ in order to
re-organize all the exponentials and write the evaluation as a standard Feynman
amplitude (as in section \ref{abelian}). It is thus natural to introduce a notion
of $\star$-product on functions of $X\in\su(2)\sim\R^3$ defined through:
\be
e^{\f{i}{2\kappa}Tr(Xg_1)}\star e^{\f{i}{2\kappa}Tr(Xg_2)}
\equiv
e^{\f{i}{2\kappa}Tr(Xg_1g_2)}.
\label{stardef}
\ee
This leads to introduce the following Fourier transform between the Lie algebra $\su(2)$ and the Lie group
$\SO(3)$:
\be
f(X)=\int dg\, e^{\f{i}{2\kappa}Tr(Xg)} \tl{f}(g).
\ee
This $\star$-product is non-commutative but still associative.
Moreover it can be seen as equivalent to the convolution product on the group:
\be
\wtl{(\phi\star\psi)}=\wtl{\phi}\circ_G\wtl{\psi},
\qquad\textrm{with}\quad
\wtl{\phi}\circ_G\wtl{\psi}(g)=\int dh\,\wtl{\phi}(gh^{-1})\wtl{\psi}(h).
\ee
We can also define the star convolution product on $\R^3$ which is dual to the
usual product of functions on the group.
\be
\wtl{(\phi \,{\circ}_{\star}\psi)}=\wtl{\phi}\wtl{\psi},
\qquad\textrm{with}\quad
\phi \,{\circ}_\star \psi (X) \equiv
\int_{\R^3}d^3X \phi(X-Y)\underset{Y}{\star}\psi(Y).
\ee
A natural set of functions to study are the Fourier transforms $f_j$
of the character $\chi_j(g)$. It is easy to derive that:
\bes
f_j \star f_k &=&\f{\delta_{jk}}{d_j}f_j, \\
f_j {\circ}_\star f_k &=& \sum_{l=|j-k|}^{j+k}f_l.
\ees
More generally, if we denote $D^j_{ab}(g)$ the matrix elements in the spin $j$ representation, we can
define:
$$
f^j_{ab}(X)=\int dg\, e^{\f{i}{2\kappa}Tr(Xg)} D^j_{ab}(g),
$$
we can check that:
\bes
f^j_{ab}\star f^k_{cd}&=&\f{\delta_{jk}}{d_j} \delta_{bc} f^j_{ad} \nn\\
f^j_{ab}\circ_\star \left(f^k_{cd}\right)^\dag&=& \sum_{l,e,f} C^{jkl}_{ace}C^{jkl}_{bdf}f^l_{ef},
\ees
where we have defined $\wtl{f^{k\,\dag}_{cd}} (g)=D^k_{cd}(-g^{-1})$ and the Clebsh-Gordan coefficient $C^{jkl}$  of the recoupling theory for representations of $\SU(2)$. This shows that the $f_j$'s project onto the $j$ representation for the $\star$-product:
$$
f_j\star f^k_{cd}=\f{\delta_{jk}}{d_j}f^j_{cd}.
$$
In particular, $f_0$ is a projector:
\be
f_0\star f^k_{cd}=\delta_{0k}f_0,
\ee
and plays the role of the usual $\delta(x)$ distribution.
In order to understand the behavior of the functions $f_j$, it is useful to look at the link between this algebra/group Fourier transform and the standard Fourier transform on $\R^3$. We can express $\SO(3)$ group elements in terms of their projections on the Lie algebra
 \be
 g=\sqrt{1-\kappa^2 |\vec{P}|^2} +i\kappa\vec{P}.\vec{\sigma}
 \ee
 A function $f$ on $\SO(3)$ can be equivalently seen as a function on the 3-ball
$B^3_\kappa=\{|\vec{P}|\leq \kappa^{-1}\}$ \footnote{ If f is continuous at the identity, it
satisfies the additional  condition
$f(\kappa^{-1} \vec{n})$ is independent of $\vec{n}$ a unit vector.
we will not impose this condition in general}.
We call this space of function $\wtl{\cal B}_\kappa(\R^3)$.
This is our Fourier space: by construction no momenta in this space takes a value larger than $\kappa^{-1}$.
Our group Fourier transform is related to the usual fourier transform by:
\be\label{Ftgroup}
f(X)=\f{\kappa^3}{\pi^2}\int_{B^3_\kappa} \f{d^3\vec{P}}{\sqrt{1-\kappa^2|\vec{P}|^2}}\,
e^{-i\vec{X}.\vec{P}} \tl{f}(P).
\ee
The functions $f_j$ can be seen to be related to the Bessel functions of the first kind. The important feature is that $f_0$ has a non-zero width. Therefore, as $f_0$ plays the role of $\delta(X)$ for the
$\star$-product, this means we have access only to a finite resolution: the $\star$-product tells us of a minimal  length scale (maximal resolution) on the spacetime ($X$ sector). This is simply due to the fact that the momentum space ($P$ sector) is bounded.

In order to describe the functional space dual to $\wtl{\cal
B}_\kappa(\R^3)$, we introduce the following kernel: \be G(X,Y) =
\int_{B^3_\kappa} \f{d^3\vec{P}}{(2\pi)^3}
e^{i(\vec{X}-\vec{Y})\cdot\vec{P}}. \ee This is a projector as we
can check that $\int d^3X\, G(X,Y)G(Y,Z)=G(X,Z)$. It is now
straightforward to show that if $f(X)$ is the group Fourier
transform of a function $\tilde{f}$ then $G(f) =f$. Moreover any
function in the image of $G$ has a Fourier transform with support
in $B^3_\kappa$, therefore the image of $L^2(\R^3)$ by this
projector, denoted ${\cal B}_\kappa(\R^3)$, is isomorphic to
$\wtl{\cal B}_\kappa(\R^3)$ by the group Fourier transform
(\ref{Ftgroup}). This Fourier transform is in fact  an isometry if
we provide $\wtl{\cal B}_\kappa(\R^3)$ and ${\cal B}_\kappa(\R^3)$
with the following norms: \bes
||\phi||_{{\cal B}_\kappa}^2 &\equiv& \int_{\R^3} \f{d^3X}{8\pi \kappa^3} (\phi\star \bar{\phi})(X), \nn\\
||\wtl{\phi}||_{\wtl{\cal B}_\kappa}^2 &\equiv&
\f{\kappa^3}{\pi^2}\int_{\wtl{\cal B}_\kappa}\f{d^3P}{\sqrt{1-\kappa^2|\vec{P}|^2}} |\wtl{\phi}(\vec{P})|^2.
\ees

With the tool of this star-product, the Feynman evaluation
\Ref{PRfeyn} reads:
\be
\ii(\Gamma,\theta)= \int\prod_v
\f{d^3X_v}{8\pi\kappa^3}\, \int\prod_e dG_e\, \Delta(\kappa m) \delta_{\kappa m_e}(G_e)\,
\prod_v
\left( \underset{e\in\pp v}{\bigstar} e^{\f{i}{2\kappa}\,Tr(X_vG_e^{\e_v(e)})}\right).
\ee
When written in terms of momenta
\footnote{The normalizations relating $g$ and $P$ are given by
\bes
 \int dg &=& \f{\kappa^3}{\pi^2} \int_{{{B}}_\kappa}\frac{d^3\v P}{\sqrt{1-\kappa^2|P|^2}},\\
 \delta_{\kappa m }(g) &=&\f{\pi}{2\kappa^2} \f{\cos \kappa m}{\sin{\kappa m}}
 \delta\left(|\vec{P}_e|^2-\f{\sin^2 \kappa m_e}{\kappa^2}\right),\\
 \Delta(\kappa m ) &=& \sin{\kappa m}.
 \ees}
\be
\ii(\Gamma,\theta)=\left(\prod_e \f{\pi\cos \kappa m_e}{2\kappa^2}\right) \int\prod_v
\f{d^3X_v}{8\pi\kappa^3}\,
\int\prod_e
\f{\kappa^3 \,d^3\vec{P}_e}{\pi^2 \sqrt{1-\kappa^2|P|^2}}\,
\delta\left(|\vec{P}_e|^2-\f{\sin^2 m_e\kappa}{\kappa^2}\right)\,
\prod_v
\left( \underset{e\in\pp v}{\bigstar} e^{{i}\,{\e_v(e)}\vec{X}_v\cdot\vec{P}_e}\right).
\ee

So far the propagator which enters this amplitude is the Hadamard propagator
$\delta\left(|\vec{P}_e|^2-\f{\sin^2 m_e\kappa}{\kappa^2}\right)$. We can write
this propagator as a proper time integral:
$$
\delta\left(|\vec{P}_e|^2-\f{\sin^2 m_e\kappa}{\kappa^2}\right)=
\int_{-\infty}^{+\infty} \f{dT}{2\pi}\,e^{iT\left(|\vec{P}_e|^2-\f{\sin^2 m_e\kappa}{\kappa^2}\right)}.
$$
If one restricts the integral to be over positive proper time $T\in\R_+$, one obtains the usual
Feynman propagator with a renormalized mass. Using this Feynman propagator, the spin foam amplitude reads
\be
\ii_F(\Gamma,\theta)= \prod_e\left(\f{\cos \kappa m_e}{4\kappa^2}\right)
\int\prod_v \f{d^3X_v}{8\pi\kappa^3}\,
\int\prod_e dg_e\,
\f{i}{|\vec{P}_e|^2-\f{\sin^2 m_e\kappa}{\kappa^2} +i\epsilon}\,
\prod_v
\left( \underset{e\in\pp v}{\bigstar} e^{{i}\,{\e_v(e)}\vec{X}_v\cdot\vec{P}_e}\right).
\label{IF}
\ee
with $\v{P}_e\equiv \v{P}(g_e)$.

In the next section, we will show that these amplitudes are truly the Feynman graph evaluations of a non-commutative field theory for a $\kappa$-deformed Poincar\'e group. Such field theory then acquires the interpretation of an effective theory for matter propagating in the 3d quantum geometry.

\sss{Non-spherical Graphs and non-trivial Braiding}
\label{braided}

Beyond the case of spherical graphs, dealing with graphs with an
surface embedding of non-trivial topology $g\ne 0$ is more subtle.
It will lead us to the notion of braided Feynman diagrams for a
non-commutative field theory.

Starting with the graph amplitude \Ref{zg1} and defining the edge
group elements $H_e=g_{t(e)}a_e^{\e(e)}g_{s(e)}^{-1}b_e^{\e(e)}$,
it is obvious that the product of such $H_e$ variables around a
vertex $v\in\Gamma$ will not generally be the identity. First, let
us notice an ambiguity in the definition of the group element
associated to an edge. Indeed the distribution $\delta_\theta$ is
invariant under conjugation
$\delta_\theta(\cdot)=\delta_\theta(k\cdot k^{-1})$, so that we
can in general choose some arbitrary variables $k_e$ and use the
group elements $G_e=k_eH_ek_e^{-1}$. To fix this ambiguity and
build rigorously the group variables $G_e$, one uses a methods
similar to the gauge-fixing techniques used in \cite{gf,PR,nc}. To
follow the procedure, it is easier to work on the dual graph to
$\Gamma$. One chooses a dual vertex $\vb_0$ of reference (an
initial face $f_0$) and a maximal tree $T$ on the dual graph. For
any dual vertex $\vb$, there exists a unique path ${\cal
P}(\vb_0\arr \vb)$ running from $\vb_0$ to $\vb$ along the tree
$T$. Then one can define the ordered product $k_\vb$ of group
elements $g_{\wb}$ associated to the dual vertices along the path
${\cal P}(\vb_0\arr \vb)$. One should moreover include all the
cycles $a_i,b_i$ that the path crosses: if the dual edge between
the consecutive dual vertices $\wb$ and $\wb'$ intersects a cycle
(i.e if the corresponding edge belongs to the cycle), then one
includes the group element corresponding to that cycle in the
ordered product $k_\vb$. Then one would define the edge group
element $G_e=k_{t(e)}H_ek_{s(e)}^{-1}$, which is simply the
ordered product of face group elements $g_f$ and cycle variables
around the edge $e$ but starting at a fixed reference face $f_0$.
Nevertheless, the (ordered) product of the $G_e$'s around all
vertices is still not trivial and the situation is more subtle
than a simple choice of origin on the dual complex.

\medskip

The moot point is at the level of crossings: when two edges of the
graph cross each other (on a 2d projection of the graph or more precisely when embedding the graph on the sphere), we have a non-trivial braiding of the edge group elements, which forces us
to associate two group elements to each edge, one at the source
vertex and one at the target vertex.

\begin{figure}[t]
\begin{center}
\psfrag{1}{$1$}
\psfrag{2}{$2$}
\psfrag{3}{$3$}
\psfrag{4}{$4$}
\psfrag{5}{$5$}
\psfrag{a}{$a$}
\psfrag{b}{$b$}
\psfrag{t1}{$\theta_1$}
\psfrag{t2}{$\theta_2$}
\psfrag{g1}{$G_A$}
\psfrag{g2}{$G_B$}
\includegraphics[width=11cm]{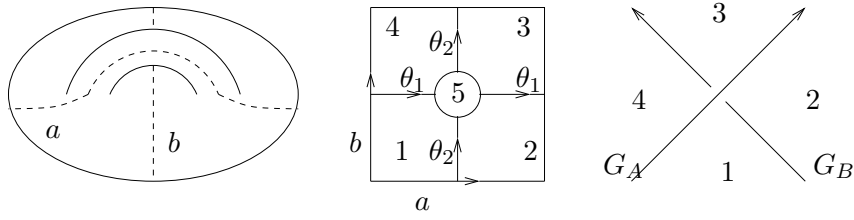}
\end{center}
\caption{Case of a crossing: we lift the crossing  by embedding it
into a handle, we triangulate the corresponding punctured torus
and we derive the braiding rule in the non-abelian Feynman
evaluation.} \label{crossing}
\end{figure}

First  $\Gamma$ being a non-spherical graph means that we
cannot draw it on a sphere without crossings. In order to compute
the amplitude we need to choose a surface on which this graph is
drawn without crossing. There is an arbitrariness on the choice of
this surface of course. Our amplitude being topological, it
doesn't matter which surface we choose and we are free
to choose the most convenient one. The surface which we use
consists of adding one handle to the sphere for each crossing: the
upper edge goes through the handle whose meridian is a cycle $a$ and
the lower edge goes below the arch of the crossing as shown in
fig.\ref{crossing}. This means that starting from a projection
of our diagram on the sphere we cut a small disk around each
crossing of the projection and glue back a punctured torus (with one hole).
Topological invariance insures that the amplitude doesn't depend
on the choice of projection of our graph.

Then let us focus on one crossing.
We choose the  simplest cell decomposition of the punctured torus
as on fig.\ref{crossing}. There is $4$ faces to which we assign
group elements $g_1,.., g_4$. In this diagram there are $8$
edges. Each edge lead to a potential constraint. $4$ of them are
mass constraints, only two (one per edge) are relevant. Among the
other 4 constraints, one can be gauged away (Lets say the one associated with the
edge bounding the face $2$ and $3$). Then the spin amplitude
\Ref{zg1} for the concerned faces and edges around the crossing
reads:
$$
\delta_{\theta_1}(g_1g_4^{-1})
\delta_{\theta_2}(g_1g_2^{-1})
\delta(g_2g_1^{-1}b)\delta(g_3g_4^{-1}b)\delta(g_4ag_1^{-1})
$$
Integrating over the cycle group elements $a,b$ leaves a single
constraint $\delta(g_1g_2^{-1}g_3g_4^{-1})$ plus the mass
constraints. We then introduce the group elements associated to the
two edges at each end:
$$
G_A^s=g_4g_1^{-1},
G_A^t=g_3g_2^{-1},
G_B^s=g_1g_2^{-1},
G_B^t=g_4g_3^{-1}.
$$
They satisfy the constraints $G_A^sG_B^s(G_A^t)^{-1}(G_B^t)^{-1}=1$.
In terms of these edge variables, the spin foam amplitude reads:
\be
\delta_{\theta_1}(G_A^s)
\delta_{\theta_2}(G_B^s)
\delta(G_A^sG_B^s(G_A^t)^{-1}(G_B^t)^{-1}) \delta(G_B^s (G_B^t)^{-1}).
\ee
The important thing to notice is that we do not have a constraint $\delta(G_A^s (G_A^t)^{-1})$.

\begin{figure}[t]
\begin{center}
\psfrag{1}{$g_1$}
\psfrag{2}{$g_2$}
\psfrag{3}{$g_3$}
\psfrag{b}{$g'_1$}
\psfrag{a}{$g'_2$}
\psfrag{V}{$\equiv \delta(g_1g_2g_3)$}
\psfrag{P}{$\equiv \delta_\theta(g_1)$}
\psfrag{B}{$\equiv \delta(g_1g_2g'_1{}^{-1}g'_2{}^{-1})\,\delta(g_2g'_2{}^{-1})$}
\includegraphics[width=3cm]{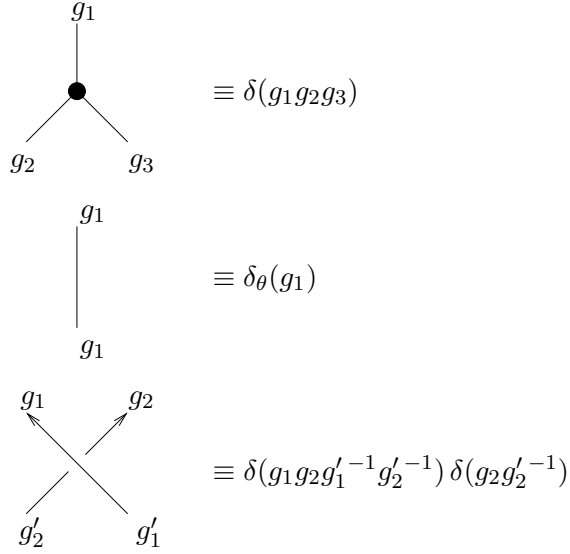}
\end{center}
\caption{Feynman rules for particles propagation in the Ponzano-Regge model.}
\label{Feynmrules1}
\end{figure}

We can now fully specify our Feynman rules. We first assign two
group elements to each edge $G_{s(e)},G_{t(e)}$. For each vertex
we put a weight $\delta(\prod_{e\supset v}G_{s(e)})$, for each
edge with no crossing we put a weight $
\delta_{\theta_e}(G_{s(e)})\delta(G_{s(e)}(G_{t(e)})^{-1})$ and
for each crossing we add $
\delta(G_{s(e')}G_{s(e)}(G_{t(e')})^{-1}(G_{t(e)})^{-1})
\delta(G_{s(e)}(G_{t(e)})^{-1}) $ where $e'$ over-crosses $e$.
This is summarized in fig.\ref{Feynmrules1}. We show in the next
section that these Feynman rules  can be derived exactly from a
non-commutative field theory based on the $\kappa$-deformed
Poincar\'e group.

These Feynman rules can be understood as a Reshetikhin-Turaev
evaluation of a colored graph constructed with the quantum group
${\cal D}(SU(2))$ (which is a $\kappa$-deformation of the
Poincar\'e group) \cite{PRII}. They involve a non-trivial braiding
factor for each crossing. We can do the same trick as previously
and promote the Hadamard amplitudes to Feynman amplitudes by
taking only the integration over positive proper time in the
propagator and we call the resulting amplitudes $\ii_F(\Gamma,\theta)$.

\section{A Non-Commutative Braided Field theory}

\subsection{Non-Commutative Field Theory as Effective Quantum Gravity}

We now introduce a non-commutative field theory based on the previous $\star$-product, and we show it leads to a sum over graphs of the particle insertion amplitudes obtained from the spin foam model. It has the natural interpretation of an effective field theory describing the dynamics of matter in quantum gravity after integration of the gravitational degrees of freedom.

Let us now consider the graph evaluations $\ii_F(\Gamma,\theta)$ given by the equation \Ref{IF}.
We consider the case where we have particle of only one
species so all masses are equal, $m_e\equiv m$. Having different masses would only require to introduce more fields and would not modify the overall picture in any way.
We define a normalized amplitude\footnote{From the spin foam point of view each amplitude correspond to
a manifold of different topology. The spin foam approach does not specify
how one should relatively weight amplitude of different
topology and we have to make a natural choice.}
$\wtl{\ii}_F(\Gamma,\theta) = \left(\f{4\kappa}{ \cos \kappa m}\right)^{|e_\Gamma|} \ii_F(\Gamma,\theta)$
where $|e_\Gamma|$ is the number of edges of $\Gamma$. Let us now consider the sum over planar and trivalent graphs:
\be
\sum_{\Gamma {\mathrm{planar, trivalent}}}
\f{\lambda^{|v_\Gamma|}}{S_\Gamma} \wtl{\ii}_F(\Gamma,\kappa m)
\ee
over trivalent planar diagrams (we will include non-planar diagrams in the
next section) where $\lambda$ is a coupling constant, $|v_\Gamma|$ is the number of vertices of
$\Gamma$ and $S_\Gamma$ is the symmetry factor of the graph.
Remarkably, this sum can be obtained from the perturbative expansion of a
 non-commutative field theory given explicitly by:
\be
S=\f{1}{8\pi\kappa^3}
\int d^3x\,\left[\f{1}{2}(\partial_i\phi \star \partial_i\phi)(x)^2 -\f{1}{2}\f{\sin^2
m\kappa}{\kappa^2} (\phi \star \phi)(x)^2 + \f{\lambda}{3!} (\phi \star \phi\star \phi)(x)
\right]
\ee
where the field $\phi$ is in $\wtl{\cal B}_\kappa(\R^3)$.
Its moment has support in the ball of radius $\kappa^{-1}$.
We can write this action in momentum space
\be
S(\phi)=
\f{1}{2}\int dg\, \left(P^2(g)- \f{\sin^2\kappa m}{\kappa^2}\right)\wtl{\phi}(g)\wtl{\phi}(g^{-1})
+\f{\lambda}{3!}\int dg_1dg_2 dg_3 \,\delta(g_1g_2g_3)\,\wtl{\phi}(g_1)\wtl{\phi}(g_2)\wtl{\phi}(g_3),
\ee
from which it is straightforward to read the Feynman rules and show our statement.

It is interesting to write down the interaction term in terms of the momenta $P(g)$. The momentum addition rule becomes non-linear, in order to preserve the condition that momenta is bounded, explicitly:
\be
\label{oplus}
\v{P}_1\oplus \v P_2 = \sqrt{1-P_2^2}\,\v P_1 +\sqrt{1-P_1^2}\, \v P_2 - \v P_1\wedge \v P_2.
\ee
Therefore the momentum conservation at the interaction vertex reads:
\be
P_1\oplus P_2 \oplus P_3=0=P_1+P_2+P_3 -\kappa(P_1\wedge P_2+ P_2\wedge P_3+P_1\wedge P_3) +{\cal O}(\kappa^2).
\ee
From this identity, it appears that the momenta is non-linearly conserved,
 and the non-conservation is stronger when the momenta are non-collinear.
 The natural interpretation is that part of the energy involved in the collision process
 is absorbed by the gravitational field, this effect prevents any energy involved in a
 collision process to be larger than the Planck energy.
 This phenomena is simply telling us that when we have a high momentum
 transfer involve in a particle process, one can no longer ignore
 gravitational effects which are going to modify how the energy is
 transferred.

The non-commutative field theory action is symmetric under a $\kappa$-deformed action of the Poincar\'e group. If we denote by $\Lambda$ the generators of Lorentz transformations and by $T_{\vec{a}}$ the generators of translations, it appears that the action of these generators on one-particle states is undeformed:
\bes
\label{1paction}
\Lambda\cdot \tl{\phi}(g) &=& \tl{\phi}(\Lambda g \Lambda^{-1})= \tl{\phi}(\Lambda \cdot P(g)), \\
T_{\vec{a}}\cdot\tl{\phi}(g) &=& e^{i\vec{P}(g)\cdot\vec{a}}\tl{\phi}(g).
\ees
Since there seems to be a certain amount of confusion in the literature on this subject, we would like to emphasize
that it is impossible to deform non-trivially the action of the Poincar\'e group on one-particle states. Indeed, it is well known that the cohomology group of the Poincar\'e group is trivial: any deformation of the Poincar\'e group which is connected to the identity (i.e. which depend continuously on a parameter $\kappa$)
can be undone by a non-linear redefinition of the algebra generators.
More precisely, let us start from any $\kappa$-deformation
(which preserve associativity or Jacobi identity)
of the Poincar\'e algebra, that is we have Poincar\'e generators $J_{\mu\nu} ,P_\mu$
with deformed commutation relations such that when $\kappa=0$ we recover the usual algebra.
Then we can always choose new generators $\tl{J}_{\mu\nu} ,\tl{P}_\mu$
which are non-linear functions the original generators and $\kappa$ such that
the new generators satisfy the undeformed Poincar\'e algebra. This choice of generators is referred as
the classical basis. This is nicely exemplified in \cite{Jerzy} for the 3+1 $\kappa$-Poincar\'e algebra
and in \cite{dsr} for the 2+1 $\kappa$-Poincar\'e algebra.
The open problem in this context is whether physics does depend on the choice of basis or not and if so which one is preferred.
It seems that the choice of basis matters, at first sight, since each basis correspond to
a choice of what should be used as an operational notion of energy\cite{ameca},
and for instance the dispersion relation is dependent on the basis choice.
Whether this is a true physical dependance or not is still however a matter of debate \cite{Jerzy2}.

It has  already been
noticed that particles in 3d quantum gravity should be described
by a Deformed/Doubly Special Relativity (DSR) \cite{PR,dsr}.
What we learn from the present analysis  is that
2+1 gravity chooses for us a preferred basis which is a classical
basis (this fact was already noticed in \cite{dsr}).
Note however that if $J_{\mu\nu},P_\mu$ form a classical basis, then
any redefinition of the form $\tl{J}_{\mu \nu}=J_{\mu\nu},\tilde{P}_\mu=P_{\mu}f({\kappa|P|})$,
with an arbitrary function $f$, still yields a classical basis.
So 2+1 gravity chooses one particular classical
basis\footnote{If we denote $p$ the standard classical basis and $P$
the one appearing in our context the relation is given by
$P^a= \f{\sin\kappa |p|}{\kappa |p|}p^a$.}. This chosen one, singled out by gravity, is consistent
with the DSR principle since it is not possible in this basis to have an energy which exceed
the Planck energy.

In a classical basis, the non-trivial deformation of the Poincar\'e group appears at the level of
multi-particle states. Indeed, the action of the Poincar\'e group on two-particle state is given by
\bes
\label{2paction}
\Lambda \cdot\tl{\phi}(P_1) \tl{\phi}(P_2)&=& \tl{\phi}(\Lambda \cdot P_1)\tl{\phi}(\Lambda \cdot P_2), \\
T_{\vec{a}}\cdot\tl{\phi}(P_1) \tl{\phi}(P_2) &=& e^{i\vec{P}_1\oplus \vec{P}_2\cdot\vec{a}}\tl{\phi}(P_1)\tl{\phi}(P_2).
\ees
The action of the Lorentz generator is undeformed\footnote{This is
different from the $\kappa$-deformation of the 3+1 Poincar\'e algebra.} however the action of the translations
is modified in a non-trivial fashion reflecting the fact that the spacetime has become non-commutative.
With these rules it is clear that our field action is $\kappa$-Poincar\'e invariant.
In algebraic language this means that the symmetry algebra is promoted to a non-cocommutative Hopf algebra. In the classical basis, the only non-trivial coproduct is\footnotemark
\be
\Delta(P_i) =\sqrt{1-P^2}\otimes P_i + P_i\otimes \sqrt{1-P^2} - \epsilon_{ijk} P^j\otimes P^k.
\ee
This shows that the effective theory describing the dynamics of second quantized particles
in the presence of gravity can be effectively described, after integration over
the gravity degrees of freedom as an explicit non-commutative field theory which respects DSR principles.
\footnotetext{This is exactly the Hopf algebra structure of the $\kappa$-deformation  of the 2+1 Poincar\'e algebra, or equivalently of the quantum double ${\cal D}(\SU(2))$, which appears in the quantization of 2+1 gravity \cite{PRII,catherine}. The infinitesimal deformation of the algebra is simply given by:
$\delta(J_i)=0,\,\delta(P_i)=\epsilon_{ijk} P^j\otimes P^k$. The $\kappa$-deformation  of the Poincar\'e algebra in 3+1 dimensions is actually different and usually involves singling out the time direction and also deforming the Lorentz generators.}

In fact we will now see by looking at the non-planar graphs
that the proper effective description is in term of a braided non-commutative field theory which
takes into account the non-trivial statistics induced by gravity.
Also, the Euclidean gravity amplitudes which we have discussed so far
are expressed in terms of Hadamard propagator and we had to extract
the Feynman propagator from them, in order to make contact with non-commutative field theory.
We will argue in the next section that the Lorentzian gravity spin foam amplitudes should
have a formulation in which gravity amplitudes are given in terms of causal propagators.

\subsection{Non-Commutativity and Braided Feynman Diagrams}

The question is now whether the non-commutative field theory introduced above reproduces
the quantum gravity amplitude in the non-planar case.
The computation of Feynman amplitudes in the planar and non-planar case differs
in the sense that in the non-planar case we have to commute the Fourier modes of the field
before doing any Wick contraction. For instance if we compute using Wick theorem the following vacuum
 expectation value, we get
\bes
\bra\wtl{\phi}(P_1)\wtl{\phi}(P_2)\wtl{\phi}(P_3)\wtl{\phi}(P_4)\ket
&=&
\bra\wtl{\phi}(P_1)\wtl{\phi}(P_2)\ket\,\bra\wtl{\phi}(P_3)\wtl{\phi}(P_4)\ket
+ \bra\wtl{\phi}(P_1)\wtl{\phi}(P_4)\ket\, \bra\wtl{\phi}(P_2)\wtl{\phi}(P_3)\ket \nn \\
&&
+\,\bra\wtl{\phi}(P_1)\wtl{\phi}(P_3)\ket\, \bra\wtl{\phi}(P_2)\wtl{\phi}(P_4)\ket.
\ees
If we draw the corresponding Feynman diagrams, putting the momenta ordered on a line
we see that the first two contractions are planar whereas the third one contain a crossing,
the crossing is due to the fact that we have to exchange $\wtl{\phi}(P_2)$ and $\wtl{\phi}(P_3)$ before
making the Wick contraction.
In order to compute the non-planar Feynman diagrams we therefore have to specify the rules
for commuting the modes $\wtl{\phi}(P)$, that is specify the {\it statistics} of our particles.
In standard commutative local field theory, we know that only two statistics are usually possible
bosons or fermions (except in $2+1$ dimensions). What about non-commutative field theory?
It seems that this is terra incognita since the usual spin statistics theorem cannot be applied
(except for some particular examples in usual non-commutative field theory with no space time non commutativity \cite{NCspst}).
We now argue  that once we have fixed the star product and the duality
between space and time there is a natural  way to specify the statistics
of our field. Let us look at the product of two identical fields:
\be
\phi\star \phi\,(x) = \int dg_1 dg_2\,
 e^{\f{i}{2\kappa} \tr(xg_1g_2)} \wtl{\phi}(g_1)\wtl{\phi}(g_2),
\ee
We can move $\tl{\phi}(g_2)$ to the left by making the following change of variables
$g_1\rightarrow g_2$ and $g_2\rightarrow g_2^{-1}g_1g_2 $, the star product reads
\be
\phi\star \phi\,(x) = \int dg_1 dg_2\,
 e^{\f{i}{2\kappa} \tr(xg_1g_2)} \wtl{\phi}(g_2)\wtl{\phi}(g_2^{-1}g_1g_2),
\ee
This suggest that the proper way to read the statistics of our non commutative field is to assume
that they satisfy the commutation relation:
\be\label{stat}
\tl{\phi}(g_1)\tl{\phi}(g_2)=
\tl{\phi}(g_2)\tl{\phi}(g_2^{-1}g_1g_2)
\ee
In our case, we can check that this commutation relation is
exactly the one coming from the braiding of two particles coupled to quantum gravity.
This braiding was computed in the spin foam model \cite{PR} and is encoded into a braiding matrix
\be
R\cdot\tl{\phi}(g_1)\tl{\phi}(g_2)=
\tl{\phi}(g_2)\tl{\phi}(g_2^{-1}g_1g_2).
\ee
This is the $R$ matrix of the $\kappa$-deformation of the Poincar\'e group \cite{schroers}.
We see that the non-trivial statistics imposed by the study of our non-commutative field theory
is related to the braiding of particles in 3 spacetime dimensions\footnote{It is often believe that this is
true only because we are studying three-dimensional field theory and this effect should disappear in higher dimension because
it is impossible to have a non-trivial statistics in $3+1$ dimensions. This relies on the fact that
in a classical spacetime the homotopy group of the configuration space of $n$ particles on the sphere
is the pure symmetric group. This belief amounts to suppose that such a theorem still holds true in the context
of a non-commutative space time. This is far from being obvious and there is no evidence supporting this hypothesis.
Moreover anybody who has dealt with non-commutative space-times knows that in a deep non-commutative regime,
the notion of dimension is not sharply defined, and our argument tends to be insensitive to dimensionality.}
Such field theory with non-trivial braided statistics are usually simply called braided non-commutative field
theory and they were first introduced in \cite{Robert}.

If one uses the non-trivial statistic (\ref{stat}), it becomes an easy exercise to check that the
partition function of our non-commutative field theory reproduces the sum over all quantum gravity
amplitudes with insertions of Feynman propagators.





\section{On the classical limit of the Turaev-Viro model}

\subsection{The many limits of the Turaev-Viro model}

We have dealt so far with the case of a non zero cosmological constant $\Lambda$.
When $\Lambda \neq 0$ the quantum gravity amplitudes are known to be given by the Turaev-Viro model
instead of the Ponzano-Regge model. We want to show in this section that the zero gravity limit of Turaev-Viro reproduces
computation of Feynman diagram with insertion of the Hadamard propagator on $\S^3$.
We also present a modification of the Turaev-Viro model which reproduces, in the no gravity limit,
 Hadamard-Feynman diagrams
on the hyperbolic space $H^3$.

If we consider the case of a non-zero cosmological constant we have three dimensionfull constants
at our disposal,the planck constant $\hbar$, the Newton constant $\kappa$ and the cosmological scale $L=1/\sqrt{\Lambda}$. We have two lengths at our disposal: the  maximal cosmological length and a minimal planck scale $l_P=\hbar \kappa$ (keep in mind that we have set the speed of light $c$ to 1). We can therefore introduce a dimensionless ratio
\be\label{kdef}
\frac{k}{\pi}  = \frac{L}{\hbar \kappa}.
\ee
$k$ is then quantized and labels the quantum group deformation parameter $q= \exp\frac{i\pi}{k}$ (see e.g \cite{kirillvolume}).  The two length scales also give rise to two mass scales.
The physical implication of these mass and length scales imply that any change of
length or mass $\Delta l$, $\Delta m$ is
bounded from above and below
\bes\label{bounds}
\hbar \kappa < &\Delta l& < \pi L,\\
\frac{\hbar}{L} < &\Delta m& < \frac{\pi}{\kappa}.
\ees
The  bounds on $\Delta l$ imply that $l$ is discrete in Planck unit and bounded from above and similarly the mass is discrete in cosmological units and bounded from above.
This is indeed what is realized in the Turaev-Viro model.
It defines a state sum model where the spins $j$ representing the geometrical information
label representations of a quantum group $SU(2)_q$ and the summation over $j$ is bounded from above
$d_j=2j+1<k$. Then the physical length is related to the spin by
\be
l= \hbar \kappa d_j.
\ee
The insertion of spinless particles in this model have been described by Barrett
in \cite{Barrett1} and are analogous to the insertion of particle in the Ponzano-Regge model.
Instead of inserting  $\Delta(\theta)\chi_j(h_\theta)$ in the state sum measure along the edge of the
Feynman graph, one puts $S_{ja}$ where $a$  is a quantum group representation label
(so $d_a<k$).
The spin $a$ is related to the physical mass by
\be
m= \frac{\hbar d_a}{L}
\ee
and $S$ is given by
\be
S_{ja} = \frac{\sin\frac{\pi}{k} d_j d_a}{\sin \frac{\pi}{k}}.
\ee
$S_{j0}$ is the quantum dimension.
In this model both lengths and masses are discretized and  satisfy the bounds (\ref{bounds}).
The amplitude is
\be\label{TVamp}
\zz_{q,M}(\Gamma_a)=N^{1-|v|}\sum_{\{j_e\}} \prod_{e\notin \Gamma} S_{j_{e}0} \prod_{e\in \Gamma}S_{j_ea_e}
\prod_t
\left\{
\begin{array}{ccc}
    j_{e_{t_{1}}} &  j_{e_{t_{2}}} &  j_{e_{t_{3}}} \\
    j_{e_{t_{4}}} &  j_{e_{t_{5}}} &  j_{e_{t_{6}}}
    \end{array}
    \right\}_q,
\ee
where we use the quantum group $6j$ symbol.
$N$ is a normalization parameter given by
\be
N= \sum_j S_{j0}^2 = \frac{k}{2\sin^2\frac{\pi}{k}}.
\ee
$|v|$ is the number of vertices of the triangulation.
One can show that this definition of the Turaev-Viro amplitude is equivalent to the
gauge fixed definition.

Starting from the Turaev-Viro model, we can take several limits of the fundamental constants
while keeping physical length and mass fixed:
\begin{itemize}
\item {\it The zero cosmological constant limit $ L \rightarrow \infty$}:\\
In this limit $ L\rightarrow \infty$ the upper bound on length and the lower
bound on mass disappear, we then recover Ponzano-Regge model studied previously where
lengths are discrete and unbounded, and masses continuous and bounded.

\item {\it The no gravity limit $\kappa \rightarrow 0$}:\\
This is the limit we are interested in. In this limit,
lengths become continuous and bounded while masses turn out discrete and unbounded.

\item {\it The semi-classical limit $\hbar \rightarrow 0$}:\\
In this limit both the mass and the lengths are continuous and bounded.
It would be interesting to understand better the resulting state sum model.

\end{itemize}

Note that in all these limits the deformation parameter becomes trivial $q \rightarrow 1$. But the simple statement $q \rightarrow 1$ is not enough to specify which limit we are considering. We also need to define how the basic quantities like masses and lengths are rescaled in the limit. More precisely, the limits are defined as the same mathematical limit $q\arr 1$, however, physically, we need to specify the units for the physical quantities such as the mass and the length: different choices of units lead to different effective theories.

In the following we work in units $\hbar=1$ and we consider the no-gravity limit of Turaev-Viro.
Let $x,y$ be two points on a 3-sphere of radius $L$ and denote by $l(x,y)$ the distance between them.
The Hadamard propagator on a sphere of radius $L$ is given by
\be
G_m(x,y) = \frac{\sin{m l(x,y)}}{\sin{l(x,y)}},
\quad \left(\Delta + m^2-\frac{1}{L^2}\right)G_m =0,
\ee
Where $\Delta$ is the Laplacian on the sphere.
If we denote by $x,y$ the corresponding points on the unit sphere $\SU(2)$ and by $a$ the half integer
such that $d_a = mL$ we see that the Hadamard propagator is just the character
$G_m(x,y) = \chi_a(xy^{-1})$.
The evaluation of Feynman diagram with insertion of the
Hadamard propagator of $\S^3$ is given by
\be\label{sphereint}
\zz(\Gamma,a) =\int \prod_v dx_v \prod_e \chi_{a_e}(x_{t(e)}x_{s(e)}^{-1})
\ee
The integrals
 over $x_v $ can be easily performed, and we are left with
 the evaluation of the spin network functional at the identity
 \be
\zz(\Gamma,a) = |\Phi_{({\Gamma}_{\Delta},a_{{e}})}(1)|^2.
\ee
We claim that this is the no gravity limit of the Turaev-Viro amplitude.
For instance if we consider the tetrahedral graph
as an example we have
\be
\zz(\Gamma,a) = \int \prod_I dx_I \prod_{I<J} \chi_{a_{IJ} } (x_Ix_J^{-1})
=\left\{
\begin{array}{ccc}
    a_{12} &  a_{13} &  a_{14} \\
    a_{34} &  a_{24} &  a_{23}
    \end{array}
    \right\}^2.
\ee
One can check that this  is actually the limit $\kappa \arr 0$ of the Turaev-Viro amplitude:
\be
\zz_q(\Gamma,a)=\frac{1}{N^3}
\sum_{\{j_I\}}
\prod_{I<J}S_{j_{IJ}a_{IJ}}
\left\{
\begin{array}{ccc}
    j_{34} &  j_{24} &  j_{23} \\
    j_{12} &  j_{13} &  j_{14}
    \end{array}
    \right\}_q^2
    =
    \left\{
\begin{array}{ccc}
    a_{12} &  a_{13} &  a_{14} \\
    a_{34} &  a_{24} &  a_{23}
    \end{array}
    \right\}_q^2.
    \ee
In general, it has been proven by Barrett \cite{Barrett2} that
$\zz_q (\Gamma,a)$ is given by the square of the quantum group evaluation
of the colored graph $(\Gamma,a)$. The no gravity limit of this evaluation therefore reproduces
the Feynman graph evaluation \Ref{sphereint}.

We are now going to show that the Turaev-Viro amplitude
\Ref{TVamp} can be written as the Feynman graph evaluation of
a non-commutative braided quantum field theory.
The classical propagator is a function on $\S^3\times \S^3$.
In a non commutative field theory $\S^3$ becomes a non commutative
(fuzzy or q-deformed) space and functions on $\S^3$ becomes operators.
If we pick an axis in $\S^3$ and denote $R$ the distance
from the north pole of $\S^3$ along this axis,
we can describe $\S^3$ as a stack of two spheres of radius $L\sin(R/L)$.
Our fuzzy $\S^3$ is then a stack of fuzzy two spheres.
It is well known \cite{fuzz} that the space of function on a fuzzy two sphere of a given radius
is given by $f_q(\S^2) = \mathrm{End}(V_I)$, where $V_I$ is here the quantum group representation space
of spin $I$.
The `radius' of this q-deformed sphere is given by the quantum dimension
$\mathrm{dim}_q(V_I)= \frac{\sin(\pi d_I/k)}{\sin(\pi/k)}$.
The space of function on the q-deformed $\S^3$ is then given as a stack of fuzzy spheres
\be
F_q(\S^3) \equiv \bigoplus_{I=0}^{(k-2)/2} \mathrm{End}(V_I).
\ee
A function is thus given by a collection of operators $f(I)\in \mathrm{End}(V_I)$ and the
normalized integral  is replaced by a q-trace
\be
\int_{\S^3_q} f = \f{1}{N} \sum_I d_I\mathrm{tr}_q(f(I)).
\ee

The propagator $G_a \in F_q(\S^3)\otimes F_q(\S^3)$  becomes a collection of operators $G_a(I,J) \in \mathrm{End}(V_I\otimes V_J)$  which are given by the Reshetikhin evaluation of the diagram.
\begin{figure}[t]
\begin{center}
\psfrag{i}{$i$}
\psfrag{j}{$j$}
\psfrag{k}{$k$}
\psfrag{l}{$l$}
\psfrag{a}{$a$}
\psfrag{V}{$\equiv {\rm Id}_i\otimes {\rm Id}_j\otimes {\rm Id}_k$}
\psfrag{P}{$\equiv G_a(i,j)$}
\psfrag{B}{$\equiv R(i,j,k,l)$}
\includegraphics[width=3cm]{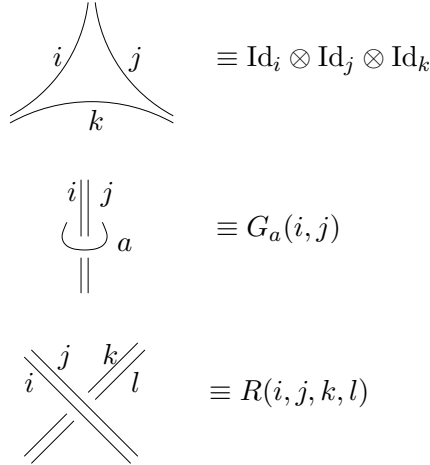}
\end{center}
\caption{Feynman rules in the Turaev-Viro model: a Feynman graph evaluation is given by the corresponding Reshetikhin evaluation of the diagram.}
\label{Feynmanrules}
\end{figure}
The Feynman rules are listed in figure \ref{Feynmanrules}. For each propagator
we insert $G_a(I,J)$, at each vertex we perform the trace and sum over $I$ and for each crossing
we insert a R-matrix.
This last factor means that, unlike the usual view on non-commutative field theory, the braiding  of the graph matters for the Feynman graph
evaluation and we are dealing with a braided non-commutative field theory\cite{Robert}. This expresses the fact that non-commutativity is naturally associated with a change of statistics of the quantum particles.

 It is easy to see by performing the sum over $I$ that our Feynman rules gives the evaluation
\be
\zz_q (\Gamma,a)= |(\Gamma,a)_q|^2,
\ee
where $(\Gamma,a)_q$ is the Reshetikhin-Turaev evaluation of
the graph $\Gamma$ colored by $a_e$.
It is less trivial to show that this is equal to the definition \Ref{TVamp}, this is proven in
\cite{Barrett2}.

\subsection{A hyperbolic quantum state sum }

We have so far considered  Feynman graph evaluation of
particle propagating in an Euclidean  space having a  positive or zero cosmological
constant.
For completeness we now consider the case where the cosmological constant is negative and the particles propagates in
the hyperbolic space $H^3$.

We claim that the corresponding state sum model is a modification
of Turaev-Viro where the deformation parameter is taken to be real
$q= \exp(-\frac{\pi}{k})$, no quantization condition holds on $k$.
The representations of $U_q(SU(2))$, $q\in\R$, are labelled by
unbounded spins $j =0,..,\infty$. The state sum model is (up to
the normalization factor, the range of summation over $j$ and the
gauge fixing)  similar to (\ref{TVamp})
 \be\label{Hypamp}
\zz_{q,M}(\Gamma, \rho)=\sum_{\{j_e\}} \prod_{e\notin T\cup
\Gamma} \mathrm{dim}_q(j_e)  \prod_{e\in \Gamma}S_{j_e\rho_e}
\prod_{e\in T} \delta_{j_e0} \prod_t \left\{
\begin{array}{ccc}
    j_{e_{t_{1}}} &  j_{e_{t_{2}}} &  j_{e_{t_{3}}} \\
    j_{e_{t_{4}}} &  j_{e_{t_{5}}} &  j_{e_{t_{6}}}
    \end{array}
    \right\}_q,
\ee
where $T$ is a maximal tree of $\Delta\setminus\Gamma$,
$\mathrm{dim}_q(j) = \frac{\sinh(\pi d_j/k)}{\sinh(\pi/k)}$
is the quantum dimension,
$\rho $ is a continuous angle $\in [0,k]$ and
\be
S_{j\rho} = \frac{\sin{\frac{\pi}{k}\rho d_j}}{\sin{\frac{\pi}{k}}}.
\ee
We now have the remarkable property that the no-gravity limit
of this amplitude is given by the Feynman graph evaluation of Hadamard propagators
in $H^3$.
The Hadamard propagator on $H^3$ is given by
\be
G_\rho(x,y) = \frac{\sin\rho l(x,y)}{\rho \sinh l(x,y)},
\ee
where $l(x,y)$ is the hyperbolic distance. The Feynman graph evaluation on $H^3$ is given by
\be
\label{Hypint}
Z(\Gamma,a) =\int \prod_v dx_v \prod_e G_{\rho_e}(x_{t(e)},x_{s(e)}).
\ee
This amplitude is an evaluation of a colored graph of $\SL(2,\C)$.
$\rho$ label simple unitary representation of $\SL(2,\C)$.
The integration at each vertex produces a simple intertwiner of $\SL(2,\C)$
according to the general picture described in \cite{fgev}, and the recombination of these
intertwiners produces the usual group evaluation of colored graph.
This evaluation is exactly the so called relativistic evaluation used as a building block
for Lorentzian gravity amplitude \cite{barrettcrane}.
For a tetrahedral graph this amplitude gives the 6j symbol of $\SL(2,\C)$

A detailed proof of this claim follows from the results of
\cite{Rochenoui,Rochemenoui}.
One can show that the state sum \Ref{Hypamp} is an evaluation
based on the non-compact quantum group $U_q(\SL(2,\C)$.
Namely $(\rho,0)$ labels the unitary simple representations of $U_q(\SL(2,\C)$.
Given such a labelling and a colored graph $(\Gamma, \rho_e)$, we consider the $U_q(\SL(2,\C)$
Reshetikhin-Turaev evaluation $|(\Gamma, \rho_e)|$, the intertwiner between the simple representations
at a vertex is constructed in \cite{Rochenoui}.
The no-gravity limit of this evaluation is then given by the
evaluation  of a graph colored by simple representation of the
group  $\SL(2,C)$. This proves the claim
once we have check that the normalization agree (the interested reader can find more details in
\cite{Rochemenoui}).

\section{Causality: Hadamard  function vs Feynman propagator}

\subsection{A brief review of Particle Propagators and Feynman Amplitudes}

If we would like to recover the (scattering) amplitudes defined by the Feynman diagrams
of the field theory, we would like to attach Feynman propagators to links of the graph
$\Gamma$ instead of the Hadamard Green function.

Let us first review the definition of the different Green functions. The basic building
blocks are the positive and negative {\it Wightman functions}.
There are solutions to the Klein-Gordon equation defined as:
\be
G^\pm(x|y)=\f{1}{(2\pi)^2}\int d^3p\,\theta(p_0)\delta(p^2-m^2)e^{\mp ip\cdot(x-y)},
\label{wightman}
\ee
and expressed in terms of correlations of the scalar field $\phi$ as:
\be
G^+(x|y)=\la0|\phi(x)\phi(y)|0\ra,
\qquad
G^-(x|y)=\la0|\phi(y)\phi(x)|0\ra=G^+(y|x)=\overline{G^+(x|y)}.
\ee
The {\it Hadamard  function} is also a solution of the Klein-Gordon equation.
Expressed as:
\be
G(x|y)=\f{1}{(2\pi)^2}\int d^3p\,\delta(p^2-m^2)e^{- ip\cdot(x-y)},
\ee
it is the sum $G=G^++G^-$.
The {\it Feynman propagator} $G_F$ satisfies the equation
$(\Box_x+m^2)G_F=-\delta^{(3)}(x-y)$ and can expressed as:
\be
iG_F(x|y)=\theta(x^0-y^0)G^+(x|y)+\theta(y^0-x^0)G^-(x|y),
\ee
\be
iG_F(x|y)=\la0|T\phi(x)\phi(y)|0\ra.
\ee
The proper time expression for $G$ and $G_F$ are:
\be
G(x|y)=\f{1}{(2\pi)^3}\int_{-\infty}^{+\infty}dT\int d^3p\,
e^{-i[p\cdot(x-y)-T(p^2-m^2)]},
\ee
\bes
G_F(x|y)&=&-i\f{1}{(2\pi)^3}\int_{0}^{+\infty}dT\int d^3p\,
e^{-i[p\cdot(x-y)-T(p^2-m^2+i\alpha)]},\label{TGF}\\
&=& \f{1}{(2\pi)^3}\int d^3p\,\f{e^{-ip\cdot(x-y)}}{p^2-m^2+i\alpha} \nn
\ees
where $\alpha>0$ is just a regularization.

Finally, one can also introduce the {\it causal Green function} $G_C$, which is a solution
of the Klein-Gordon equation. Its integral expression is:
$$
G_C(x|y)=\f{-i}{(2\pi)^2}\int d^3p\,\epsilon(p_0)\delta(p^2-m^2)e^{\mp ip\cdot(x-y)},
$$
and correspond to the correlation:
$$
iG_C(x|y)=\la0|[\phi(x)\phi(y)]|0\ra=G^+(x|y)-G^-(x|y).
$$

\medskip

Now let us consider a Feynman diagram i.e a graph $\Gamma$ with Feynman propagators living
on its edges. The evaluation of the diagram is the amplitude:
\be
I_\Gamma=\int \prod_v d\vec{x}_v\,\prod_e G^{(F)}_{m_e}(\vec{x}_{t(e)}-\vec{x}_{s(e)}).
\ee
For a given set of positions $\{\vec{x}_v\}$ and thus a particular (causal) ordering of
the vertices, the product of Feynman propagators is simply a product of Wightman functions:
\be
\prod_eG^{(F)}_{m_e}(\vec{x}_{t(e)}-\vec{x}_{s(e)})
=\prod_e G^{\epsilon_e}_{m_e}(\vec{x}_{t(e)}-\vec{x}_{s(e)}),
\ee
where the $\epsilon_e$'s record the ordering of the vertices i.e the future/past
orientation of each edge.
This way, it may seem natural to introduce Wightman diagrams, with a causal structure
$\{\epsilon_e\}$ imposed independently from the true ordering of the $x$'s, whose
evaluation will be:
\bes
I^{\{\epsilon_e\}}_\Gamma &=&
\int \prod_v d\vec{x}_v\,\prod_e G^{\epsilon_e}_{m_e}(\vec{x}_{t(e)}-\vec{x}_{s(e)})\nn\\
&=&
\int\prod_ed^3\vec{p}_e\int\prod_vd^3\vec{x}_v\,
\theta(\epsilon_ep^0_e)\delta(p_e^2-m_e^2)e^{-i\vec{p}_e\cdot(\vec{x}_{t(e)}-\vec{x}_{s(e)})}\nn\\
&=&
\int\prod_ed^3\vec{p}_e\,\theta(\epsilon_ep^0_e)\delta(p_e^2-m_e^2)
\prod_v\delta\left(
\sum_{e|v=t(e)}\vec{p}_e-\sum_{e|v=s(e)}\vec{p}_e
\right).
\ees
Because of the constraint imposing momentum conservation at the vertices, one  can solve
these constraints as explained previously and pass to the dual formulation of these amplitudes:
they can be easily reproduced by a spin foam model.

In order to do the same with the Feynman amplitude, one should get rid of the explicit use
of the causal ordering. The solution is to use the proper time formula \Ref{TGF} for $G^F$
which leads to:
\be
I_\Gamma=\int\prod_ed^3\vec{p}_e\,\int_0^{+\infty}dT_e\,
e^{iT_e(p_e^2-m_e^2+i\alpha)}
\prod_v\delta\left(
\sum_{e|v=t(e)}\vec{p}_e-\sum_{e|v=s(e)}\vec{p}_e
\right).
\ee
Now, as this includes the momenta conservation constraints, this amplitude can be re-casted in
terms of the dual (face) variables and thus easily written in terms of spin foam amplitudes.

\medskip

In the Euclidean context considered up to now, we ave been working with the Hadamard
propagators. In order to properly recover quantum field theory amplitudes, one needs
to construct (particle) observables which would reproduce the Wightman propagators and
the Feynman propagators. These notions being closely related to causality are more likely
to be properly implemented in a Lorentzian model only.
The Lorentzian analog of the Ponzano-Regge model has been considered in
\cite{LorPR} and is known to be based on the group $\SU(1,1)$.
 Therefore, we will study in
the following how to write a (well-defined) Lorentzian spin foam model based
on $\SU(1,1)$, its abelian limit as a spin foam model over the abelian group
$M_3$ (Minkowski space) and causality oriented particle insertions.
Discussion of implementation of causality in spin foam models can be found in \cite{oriti}.

An interesting feature of the Lorentz group $\SU(1,1)$ is that there already exists a split
between some so-called future oriented timelike representations and past oriented timelike
representations. Moreover, the corresponding characters is then  $\exp(\mp i d_j\theta)$
instead of $\sin(d_j\theta)$: their abelian limit will directly be the Wightman functions
instead of  the Hadamard functions. As we will see below, there is also a very natural
implementation of the Feynman propagator.

\subsection{The abelian limit of the $\SU(1,1)$ characters}

$\SU(1,1)$ has two Cartan subgroups.
The first one corresponds the (space) rotation and its elements are of the form
($0\le \theta\le 2\pi$):
$$
h_\theta=\left(
\begin{array}{cc}
\cos\theta &\sin\theta \\
-\sin\theta&\cos\theta
\end{array}
\right),
$$
while the second one is made of the boosts ($t\in\R$)
$$
\pm a_t=\pm\left(
\begin{array}{cc}
e^t & 0 \\
0 & e^{-t}
\end{array}
\right).
$$
Then the (unitary) representations of $\SU(1,1)$ are also of two kinds.
We first have the continuous series of representations, labelled by $s\in\R$. They correspond
to the (co-adjoint) orbits defined by the one-sheet hyperboloid in Kirillov's theory,
The characters are:
$$
\chi_s(h_\theta)=0
\qquad
\chi_s(\pm a_t)=\f{\cos st}{|\sinh t|}.
$$
Then we have two discrete series of representations -positive and negative series-
whose corresponding orbits are the upper and lower hyperboloid. We label them by an integer
$n\ge 1$. The characters are:
$$
\chi_n^\pm(h_\theta)=\f{\mp e^{\pm i(n-1)\theta}}{2i\sin\theta}
\qquad
\chi_n^\pm(a_t)=\f{e^{-(n-1)|t|}}{2|\sinh t|}
\qquad
\chi_n^\pm(-a_t)=(-1)^n\f{e^{-(n-1)|t|}}{2|\sinh t|}.
$$

\medskip

Now let us consider the abelian/classical/flat limit of the characters for the (positive)
discrete representations. We take the angles $\theta=m\kappa$ (or $t=m\kappa$) going to zero
while we take the representation labels $(n-1)=l/\kappa$ to $\infty$ by
taking the limit $\kappa\arr 0$.
In this abelian limit, group elements of $\SU(1,1)$ are mapped into Lie algebra vectors of
$\su(1,1)\sim M_{(3)}$: rotations correspond to time-like vectors and boosts to
space-like vectors. The limit of the character reads:
\be
\chi_n^+(h_\theta)\arr \tl{\chi}_l(x)=\f{ -e^{il |x|}}{2i |x|}
\textrm{ for } \vec{x} \textrm{ timelike},
\ee
and
\be
\chi_n^+(a_t)\arr \tl{\chi}_l(x)=\f{e^{- l |x|}}{2 |x|}
\textrm{ for } \vec{x} \textrm{ spacelike}.
\ee
Then from the point of view of the abelian limit of the Lorentzian model, it is natural
to consider Feynman evaluation with propagators given by $G_m(\vec{x})=\tl{\chi}_m(x)$.
Using the Kirillov formula:
\be
\forall \vec{x}\in M_3,\,
G_m(\vec{x})=m\int_{{\cal H}_+}d^2\vec{n} \, e^{im\vec{x}\cdot\vec{n}}
=\f{1}{m}\int_{{\cal C}_+}d^3\vec{p}\,\delta(|p|-m)e^{i\vec{x}\cdot\vec{p}},
\ee
where ${\cal H}_+$ is the future light cone and ${\cal C}_+$ the future cone.
Comparing this expression with the definition \Ref{wightman}, it is obvious that
the propagator defined as the abelian limit of the character $\chi_n^+$ is the positive
Wightman (Green) function. Similarly, if we would assign a negative representation
$n^-$ to an edge of the graph, we will put assign the negative Wightman function to that link.
Finally, we could assign the representation $n^+\oplus n^-$ to the edge, the corresponding
character would be the sum $\chi^+_n+\chi_n^-$ which would lead to the usual Hadamard (Green)
function.

\subsection{Constructing a Lorentzian spin foam model}

Let us start by constructing an abelian Lorentzian spin foam model, with observables
for particle insertion which exactly reproduce the Feynman amplitudes. Then we would like
to write a Lorentzian version of the Ponzano-Regge model, based on  $\SU(1,1)$ whose
abelian limit (corresponding to the Newton constant going to 0) reduces to that model.

Formally, we can define the abelian Lorentzian spin foam exactly the same way than
the Euclidean one:
\be
Z=\prod_\eb\int_{M_3}d^3\vec{u}_\eb\,\prod_\fb\delta(\vec{u}_\fb),
\ee
where we now label the (dual) edges by vectors in the 3-dimensional Minkowski space $M_3$.
To insert a particle on an edge $e$, we identify the momentum $\vec{p}_e$ with the
geometric variable $\vec{u}_\fb$ living on the dual plaquette. Then:
\begin{itemize}
\item to insert the Hadamard  function: one replaces the constraint
$\delta(\vec{u}_e)$ by $$\delta(|\vec{u}_e|^2-m^2).$$
\item to insert a Wightman function, one splits the $\delta(|\vec{u}_e|^2-m^2)$
in two depending on the time orientation
 and uses $$\theta(\epsilon u_e^0)\,\delta(|\vec{u}_\fb|^2-m^2).$$
\item to insert the Feynman propagator, one splits $\delta(|\vec{u}_e|^2-m_e^2)$
in two in the proper time formulation and uses
$$\int_0^{+\infty}dT_ee^{iT_e(|\vec{u}_e|^2-m_e^2+i\alpha)},$$
where $\alpha$ is just a regularization.
\end{itemize}


\medskip

Now we would like to introduce the corresponding propagators in the non-abelian case, i.e. the observables describing particles in the Lorentzian spin foam model. Technically, our aim is to identify the quantity "$\Delta(\theta)\delta_\theta(g)$" which we should substitute to $\delta(g)$ in the spin foam model. Our main criteria is to check that the non-abelian propagators have the right expected behavior in the no-gravity limit $\kappa\arr 0$.

Let us start by constructing the Hadamard propagator. The non-abelian counterpart of
$\delta(|\vec{u}_e|^2-m^2)$ is to fix the group element $g_e$ in the same conjugacy class than the Cartan element $h_\theta$. Thus we introduce the distribution $\delta_\theta(g)$ defined by:
\be
\int_{\SU(1,1)}dg\,\delta_\theta(g)f(g)=\int_{\SU(1,1)/\U(1)}du\,f(uh_\theta u^{-1}),
\,\forall f.
\ee
Let us insist at this point that $h_\theta$, $\theta\in ]0,\pi]$ is not conjugated to $h_{-\theta}$. Then as $\SU(1,1)/\U(1)\sim {\cal H}_+\cup{\cal H}_-$ is the (disjoint) union of the upper and lower mass-shell hyperboloid, we can actually write:
$$
\int_{\SU(1,1)}dg\,\delta_{\theta>0}(g)f(g)=\int_{{\cal H}_+}du\,f(uh_\theta u^{-1}),\quad
\int_{\SU(1,1)}dg\,\delta_{-\theta}(g)f(g)=\int_{{\cal H}_-}du\,f(uh_\theta u^{-1}).
$$
It is easy to check that \be
\delta_\theta(g)=\sum_{n,\epsilon}\chi_n^{-\epsilon}(h_\theta)\chi_n^\epsilon(g),
\ee since the character of the continuous representations vanish
on the $\U(1)$ Cartan elements $h_\theta$. The Hadamard
propagator is then defined as: \be
\delta_\theta^{Hadamard}(g)\,\equiv\,\delta_\theta(g)+\delta_{-\theta}(g).
\ee In the abelian limit, $\theta\in[0,\pi]$ will be put it
$\kappa$ units, $\theta=m\kappa$,  with $m$ the (renormalized)
mass. Notice that $\chi_n^\pm(-\theta)=\chi_n^\mp(\theta)$, so
that we can write: \be
\delta_\theta^{(H)}(g)=\sum_{n,\epsilon,\epsilon'}\chi_n^\epsilon(\theta)\chi_n^{\epsilon'}(g).
\label{Hadamard} \ee

Then the Wightman propagators only impose $p^0>0$ or $p^0<0$.
This is exactly whether $u\in{\cal H}_+$ or $u\in{\cal H}_-$, i.e.
distinguishing $\theta>0$ from $\theta<0$. Thus we define the
non-abelian Wightman propagators for $\theta\in]0,\pi]$: \be
\delta_\theta^{(W+)}(g)\,\equiv\,\delta_\theta(g), \qquad
\delta_\theta^{(W-)}(g)\,\equiv\,\delta_{-\theta}(g). \ee

The hard part is to define the non-abelian equivalent of the
Feynman propagator.  Looking at the proper time expression
\Ref{TGF} of the propagators, we can interpret \Ref{Hadamard} as
the equivalent expression of the Hadamard Green function in the
non-abelian case. Then the Feynman propagator should be a
particular splitting of the sum over $\epsilon$'s. To decide, let
us look at the abelian limit of
$\sum_n\chi_n^\epsilon(\theta)\chi_n^{\epsilon'}(g)$. Using the
explicit formula for the characters, it is easy to get
($|\vec{p}|\in\R_+$):
\be
\sum_n\Delta(\theta)\chi_n^\epsilon(\theta)\chi_n^{\epsilon'}(g)
\underset{\kappa\arr0}\sim \int_0^{+\infty}dl\,
\f{-\epsilon\epsilon'}{4}\f{sg(p_0)}{|\vec{p}|}
e^{il(\epsilon'sg(p_0)|\vec{p}|+\epsilon m)},
\label{limit1}
\ee
when $g=\cos\phi+isg(p_0)\sin\phi\hat{u}.\vec{\sigma}$,
$\phi\in[0,\pi]$, $\hat{u}$ normalized positive timelike vector,
is conjugated to the compact Cartan group. Here we have defined
$\kappa\vec{p}=sg(p_0)\sin\phi\hat{u}$ for $\kappa\arr 0$. If $g$
is conjugated to the non-compact Cartan group, it corresponds to a
spacelike momentum and the limit is (still $|\vec{p}|\in\R_+$):
$$
\sum_n\Delta(\theta)\chi_n^\epsilon(\theta)\chi_n^{\epsilon'}(g)
\underset{\kappa\arr0}\sim \int_0^{+\infty}dl\,
\f{-\epsilon}{4i}\f{1}{|\vec{p}|} e^{-l(|\vec{p}|-i\epsilon m)}.
$$
Inserting a regulator $\alpha>0$ in the expression \Ref{limit1}, we get:
\bes
\sum_ne^{-(n-1)\alpha}\Delta(\theta)\chi_n^\epsilon(\theta)\chi_n^{\epsilon'}(g)
&\underset{\kappa\arr0}\sim &
\int_0^{+\infty}dl\, \f{-\epsilon\epsilon'}{4}\f{s}{|\vec{p}|}
e^{il(\epsilon'sg(p_0)|\vec{p}|+\epsilon m+i\alpha)},\nn\\
&\underset{\kappa\arr0}\sim&
-\f{\epsilon\epsilon'}{4}\f{s}{|\vec{p}|}\f{i}{\epsilon'
s|\vec{p}|+\epsilon m+i\alpha},
\ees
where we note $s=sg(p_0)$.

A priori, one would say that combining these expression, one can
obtain  all the different propagators of quantum field theory with
different choices and signs of the poles. It is actually true
except for the case of the Feynman propagator whose case is
slightly more subtle.

Indeed, we can combine
$$
\f{1}{s|p|-m+i\alpha}+\f{1}{s|p|+m+i\alpha}=2s|p|\,\f{1}{|p|^2-m^2+is\alpha},
$$
which would give the Feynman propagator if $s\alpha>0$.  Therefore
the sign of the regulator would need to depend on the sign $s$ of
the group element $g$. The easiest way out is to introduce a group
element $|g|$ defined by:
$$
g=uh_\phi u^{-1}=\cos\phi Id+i\sin\phi \hat{u}.\vec{\sigma}
\,\Rightarrow\,
|g|=uh_{|\phi|} u^{-1}=\cos\phi Id+i|\sin\phi| \hat{u}.\vec{\sigma},
$$
so that always $s(|g|)=+1$. Then it is straightforward  to check
that a good choice of Feynman propagator is:
 \be
 \delta^{(F)}_\theta(g)
\equiv 2i\sum_n
e^{-(n-1)\alpha}\Delta(\theta)\chi^-_n(\theta)\chi_n(|g|) \equiv
2i\sum_n\chi_n(|g|)
(\Delta(\theta-i\alpha)\chi^-_n(\theta-i\alpha)).
 \ee
since its classical limit is the expected $1/(|p|^2-m^2+i\alpha)$.


\section*{Conclusion}
In this paper we have shown how the quantum gravity amplitudes
for particles coupled to three dimensional gravity proposed in \cite{PR}
are related to the usual amplitude in the no-gravity limit. More surprisingly, it turns out that
these amplitudes can be understood as coming from a non-commutative braided quantum field theory.

We have seen however that the natural quantum gravity amplitudes amount to insert Hadamard propagators, then we have discussed how the insertion of a Feynman propagator can be naturally implemented in the spin foam model
at least in the Lorentzian case.
What would need to be done in this context is to show that this procedure is really well-defined in the
spin foam context, namely that our proposal for introducing the Feynman propagator in the
Lorentzian Ponzano-Regge
model respect the topological invariance of the theory and leads to triangulation
independent amplitudes.
Also it would be interesting to understand more generally the definition
of the Feynman propagator in the case of non-zero cosmological constant involving quantum group.

It is not totally surprising that the effective field theory describing the dynamics of particles
coupled to three space-time dimensions is given in term of a non-commutative braided quantum field theory.
Indeed it is known that particles in 3d gravity respect the principle of Doubly special relativity
and that the corresponding algebra of symmetry is a $\kappa$-deformation of the Poincar\'e group.
Quantum field theories based on $\kappa$-deformed Poincar\'e group have been already considered in the literature
\cite{kappafield}. Usually there is an ambigu\"\i ty in the definition of the spacetime coordinates
which leads to different proposals, none of which are physically motivated.
Our analysis  gives a unique prescription and solves this ambigu\"\i ty. It seems that the explicit form of the
action and star product which we found has never been proposed.
An interesting perspective appear if one remark that the structure of the effective field
theory which arise here possess some strong similarity with the Group field theories
that appears in the spin foam quantization of gravity \cite{GFT}, it is tempting to
conjecture that this similarity is not purely accidental and that the structure which is unravelled here
should generally appear in the spin foam context.

This questions becomes more interesting if we think of 2+1 gravity as a toy model
for 3+1 gravity and hope to learn something about the effective dynamics of
matter coupled to quantum gravity.
This hope is substantiated by the recent work \cite{artem} which shows a reformulation of 3+1 gravity
in terms of a perturbed $BF$ theory where matter coupling to gravity is very similar to the 2+1 gravity case.
Moreover, it seems that the star product that we have introduced should be generalized to a 3+1 quantum space-time. A subtle point is that star products in the field theory action would then also explicitly involve polynomial terms in the coordinates (or equivalently derivations in the momenta), which are usually ignored in standard analysis of possible non-commutative field theories as effective quantum gravity theories. This specific structure allows the field theory to be actually invariant under the deformation of the Poincar\'e group.

\section*{Acknowledgements}
We would like to thank David Louapre for his active participation in the early stages
of this work.
\appendix

\section{Examples of the duality transform of Feynman diagrams}

\subsection{Example of $\Gamma\harr{\cal S}^2$}

Let us consider the tetranet embedded in ${\cal S}^2$.
We have six edges thus six (momentum) variables $\vec{a},\vec{b},\vec{c},\vec{d},
\vec{e},\vec{f}$,
and four vertices resulting in four constraints:
\be
\left\{
\begin{array}{c}
a+b+c=0 \\
a-d+e=0 \\
c+d+f=0 \\
-b+e+f=0
\end{array}
\right.
\ee
\begin{figure}[t]
\begin{center}
\psfrag{a}{$a$}
\psfrag{b}{$b$}
\psfrag{c}{$c$}
\psfrag{d}{$d$}
\psfrag{e}{$e$}
\psfrag{f}{$f$}
\includegraphics[width=3cm]{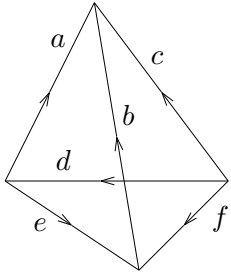}
\end{center}
\caption{Tetrahedral net as the skeleton of a 2-sphere.}
\label{tetranet}
\end{figure}
The rank of this system is 3, so that we need at least $6-3$ variables to parametrize
its solutions. A convenient parametrization is given by assigning a new variable to each
face, thus four variables $\alpha,\beta,\gamma,\delta$:
\be
a=\alpha-\delta,\quad
b=\beta-\alpha,\quad
c=\delta-\beta,\quad
d=\gamma-\delta,\quad
e=\gamma-\alpha,\quad
f=\beta-\gamma.
\ee
It is obvious that we can translate all face variables by a constant vector $\vec{k}$
and it wouldn't change anything. So, at the end of the day, we truly have 3 parameters
to describe the space of solutions, which is consistent.

More formally, one can introduce conjugate variables to the face variables
$\vec{\alpha},\vec{\beta},...$. Let us note them $\vec{v}_\alpha,...$. Then
$\vec{C}=\vec{v}_\alpha+\vec{v}_\beta+\vec{v}_\gamma+\vec{v}_\delta$ generates
simultaneous translations on $\vec{\alpha},...$, so that we can see the residual invariance
as imposed by the constraint $\vec{C}=0$. This can be understood as the closure constraints
for the tetrahedron assuming that the vectors $\vec{v}_\alpha,...$ are the normal vectors to
the faces of the tetrahedron.

\subsection{Example of $\Gamma\harr T^2$}

One can also embed the tetranet in the torus $T^2$.
\begin{figure}[t]
\begin{center}
\psfrag{a}{$a$}
\psfrag{b}{$b$}
\psfrag{c}{$c$}
\psfrag{d}{$d$}
\psfrag{e}{$e$}
\psfrag{f}{$f$}
\psfrag{A}{$\alpha$}
\psfrag{B}{$\beta$}
\includegraphics[width=3.5cm]{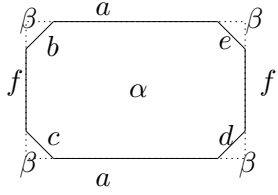}
\end{center}
\caption{Tetrahedral net as the skeleton of a 2-torus.}
\end{figure}
As the graph is the same as in the previous case, we have the same variables with
the same constraints. The difference is that we now have only two faces
$\vec{\alpha},\vec{\beta}$. This is not enough variables to parametrize the
space of solutions. We need two extra vectors, which  will be associated to the
two (non-contractible) cycles of the torus:
$$
\left\{
\begin{array}{ccc}
a&=&{\cal C}_1 \\
b&=& (\alpha-\beta)-{\cal C}_1\\
c&=& (\beta-\alpha)\\
d&=& (\alpha-\beta)-{\cal C}_2\\
e&=& (\alpha-\beta)-{\cal C}_1-{\cal C}_2\\
f&=& {\cal C}_2
\end{array}
\right.
$$
The chosen cycles, as sequences of edges, are $a,-b,-e$ and $-d,-e,f$.
Let us also notice that once again we can translate $\vec{\alpha},\vec{\beta}$
simultaneously by any vector without changing anything. This corresponds to the closure
of the faces.

\subsection{Example of $\Gamma\harr {\cal S}^2\harr {\cal S}^3$}

Let us consider the 3d manifold ${\cal S}^3$ and triangulate it using two tetrahedra.
The skeleton of the resulting triangulation $\Delta$ is simply a tetranet, which we take
as on fig.\ref{tetranet}. Let us then consider the Feynman diagram $\Gamma$ made of the edges
$a,b,c,d,e$. The graph has four vertices and the corresponding equations are:
\be
b=e,\quad
c+d=0,\quad
a+b+c=0,\quad
a+e=d.
\label{eqn1}
\ee
To parametrize the space of solutions, we introduce embed $\Gamma$ in the tetranet  considered
as a triangulation of ${\cal S}^2$. The tetranet has four faces for each of which
we introduce a new variable: $u_1$ for $abe$, $u_2$ for $acd$, $u_3$ for $bcf$ and $u_4$
for $def$. Then we can express the original momentum variables $a,..,e$ in terms of these
face variables:
$$
a=u_1-u_2, \quad
b=u_3-u_1, \quad
c=u_2-u_3, \quad
d=u_4-u_2,\quad
e=u_4-u_1.
$$
For this to be a solution of the constraints \Ref{eqn1}, it is necessary and sufficient to
impose $f=u_3-u_4=0$. And we finally recover the formula \Ref{ig}
for the Feynman amplitude.



\section{Feynman graph with Spinning particles}

\subsection{Overview of the spinning particle}

In \cite{PR} Ponzano-Regge amplitudes for spinning and spinless particles have been proposed.
In the main text we have shown that the spinless Ponzano-Regge amplitudes  reproduces in the no
gravity limit the usual Hadamard-Feynman
amplitudes. In this appendix we show that the same is true for the spinning amplitudes.
Let us present a quick review of the properties and propagator for a massive spinning
particle/field in three dimensions. Let us start by the Euclidean case, keeping that
all the considerations could be straightforwardly translated to the Minkowskian case.

Let us look at a one-particle state $\psi_{p,\sigma}$ where $p$ denotes the (3-)momentum
and $\sigma$ the other degrees of freedom:
$$
P^{\mu}\psi_{p,\sigma}=p^\mu\psi_{p,\sigma}.
$$
Considering the action of Lorentz transformations $U(\Lambda)$ on the translation generators,
$U(\Lambda)$ maps a state of momentum $p$ onto a state of momentum $\Lambda p$:
$$
U(\Lambda)\psi_{p,\sigma}=
\sum_{\sigma'}C_{\sigma',\sigma}(\Lambda,p)\psi_{\Lambda p,\sigma'}.
$$
Let now us choose a reference momentum $k^\mu$ and define states $\psi_{p,\sigma}$
from the states $\psi_{k,\sigma}$:
$$
\psi_{p,\sigma}=N(p)U(s(p))\psi_{k,\sigma},
$$
where $N(p)$ is a normalization factor and $s(p)$ maps $k$ to $p$.
For a massive particle, $k$ is usually taken as $p^{(0)}=(1,0,0)$.
The little group of Lorentz transformations $W$ leaving $k$ invariant is the rotation group
$\U(1)$.
So $s(p)$ is a section\footnote{We choose
\be
s(m\cosh\theta,m\sinh\theta\cos\phi,m\sinh\theta\sin\phi)=
e^{i\phi\sigma_3} e^{\theta \sigma_2}.
\ee} from $\SU(2)/\U(1)$ to $\SU(2)$ such that:
$$
s(p).(1,0,0)=\f{1}{m}p=\left(\sqrt{1-\f{|\vec{p}|^2}{m^2}},\f{1}{m}\vec{p}\right).
$$
Then the action of a Lorentz transformation reads:
$$
U(\Lambda)\psi_{p,\sigma}=N(p)U(s(\Lambda p))U(W(\Lambda,p))\psi_{k,\sigma},
$$
with $W(\Lambda,p)=s(\Lambda p)^{-1}\Lambda s(p)$ living in the little group $\U(1)$.
Note that $W(R,p^{(0)})=R$ for rotations and $W(B,p^{(0)})=Id$ for pure boosts.
From this, it is straightforward to check that postulating the action of the little group:
$$
U(W)\psi_{k,\sigma}=\sum_{\sigma'}D_{\sigma',\sigma}(W)\psi_{k,\sigma'},
$$
we get the action of an arbitrary Lorentz transformations:
$$
U(\Lambda)\psi_{p,\sigma}=
\left(\f{N(p)}{N(\Lambda p)}\right)
\sum_{\sigma'}D_{\sigma',\sigma}(W(\Lambda,p))\psi_{\Lambda p,\sigma'}.
$$
So that instead of giving all the coefficients $C_{\sigma',\sigma}(\Lambda)$,
it is enough to define the coefficient $D_{\sigma',\sigma}(W)$: from a representation
of the little group $\U(1)$, we induce a representation of the Lorentz group $\SU(2)$.
The usual normalization is $N(p)=\sqrt{k_0/p_0}$ so that the scalar product
is normalized
$\la\psi_{p,\sigma},\psi_{p,\sigma}\ra=
\delta_{\sigma\sigma'}\delta^{(2)}(\vec{p}-\vec{p'})$.
The {\it spin} $s$ is the choice of a irreducible representation of the little group $\U(1)$,
which defines the coefficient $D_{\sigma',\sigma}$, and $\sigma$ is the angular momentum.

All this can be directly translated to a field theory. Let us assume that the field $\psi$
has components $\psi_l$ labelled by $l$ and is a representation of the Poincar\'e group.
More precisely, we split the field into creation and annihilation parts
$\psi_l=\psi_l^++\psi_l^-$:
$$
\psi_l^+=\sum_{\sigma}\int d^2\vec{p}\,u_l(x,\vec{p},\sigma)a(\vec{p},\sigma),
\qquad
\psi_l^-=\sum_{\sigma}\int d^2\vec{p}\,v_l(x,\vec{p},\sigma)a^\dagger(\vec{p},\sigma).
$$
Then the action of a Poincar\'e transformation reads:
$$
U(\Lambda,a)\psi_l^{\pm}U(\Lambda,a)^{-1}=
\sum_{l'}D^I_{ll'}(\Lambda^{-1})\psi_{l'}^{\pm}(\Lambda x +a),
$$
where we have chosen a representation $I$ os the Lorentz group $\SU(2)$.
Interactions for such a field can be written as:
$$
V(x)=g_{l_1\dots l_Nl^\prime_1\dots l^\prime_M}
\psi^-_{l^\prime_1}(x)\dots\psi^-_{l^\prime_M}(x)
\psi^+_{l_1}(x)\dots\psi^+_{l_N}(x),
$$
where $g_{l_1\dots}$ is an intertwiner for the Lorentz group (in our case where the Lorentz
group is $\SU(2)$, it is a Clebsh-Gordan coefficient), so that the interaction is a scalar
for the Lorentz group.
Then the coefficients $u_l$ (and $v_l$) defining the field satisfy the following constraints:
$$
u_l(x,\vec{p},\sigma)=(2\pi)^{-3/2}e^{ip\cdot x}u_l(\vec{p},\sigma),
$$
$$
\sum_{\sigma'}u_{l'}(\vec{(\Lambda p)},\sigma') D^s_{\sigma'\sigma}(W(\Lambda,p))=
\sqrt{\f{p^0}{(\Lambda p)^0}}\sum_l D_{l'l}(\Lambda)u_l(\vec{p},\sigma).
$$
In order to understand the structure of this constraint, let us split it
in the two cases when $\Lambda$ is a pure boost or a rotation.
$\Lambda$ boost, with $W(\Lambda,p)=1$, allows to map the reference
momentum $^{(2)}\vec{p}=0$ to an arbitrary momentum:
$$
u_{l'}(p,\sigma)=\sqrt{\f{m}{p^0}}\sum_lD_{l'l}(s(p))u_l(0,\sigma),
$$
so that we go from $u_l(0,\sigma)$ to $u_l(p,\sigma)$ simply by a boost. Then for a
rotation $\Lambda=R$, with $W(R,p)=R$ and ${}^{(2)}\vec{p}=\vec{(Rp)}=0$,
we have:
$$
\sum_{\sigma'}u_{l'}(0,\sigma') D^s_{\sigma'\sigma}(R)=
\sum_l D^I_{l'l}(R)u_l(0,\sigma),
$$
which states that the coefficient $u_l(0,\sigma)$ can be understood as an intertwiner
between the representation of $\U(1)$ defined by the spin $s$ and the representation of
the little group $\U(1)$ induced by the representation $I$ of the Lorentz group $\SU(2)$.
So that the field definitively lives in the representation of spin $s$ of the little group.
In our Euclidean 3d context, $I$ is a representation of $\SU(2)$ and $s$, the spin, a
representation of $U(1)$ such that $s\le I$ (i.e $s\harr I$). There is only one value
of $\sigma$ which is $s$ and $D^s(R(\theta))=\exp(is\theta)$ for a rotation of angle $\theta$.
Then the unique solution to the constraints is $u_l(0)=\la l|s\ra=\delta_{ls}$.

Finally, the causal propagator is defined as the commutator of the field $\psi$ with itself.
Assuming that $v_l=u_l^\dagger$, we have:
$$
[\psi(x),\psi(y)]=\f{1}{(2\pi)^3}\int d^2\vec{p}\,\Delta_{ll'}(p)e^{ip\cdot(x-y)},
\qquad\textrm{with}\quad
\Delta_{ll'}(p)=\sum_\sigma u_l(\vec{p},\sigma)u^\dagger_{l'}(\vec{p},\sigma).
$$

To sum up, the propagator $\Delta_{ll'}^{I}(p)$ of a field of spin $s$
must satisfy the following properties:
\begin{itemize}
\item $\Delta(\Lambda p\Lambda^{-1})=\Lambda \Delta(p)\Lambda^{-1}$.
\item $\Delta(p^{(0)})$ is the projector $|s\ra\la s|$ to the right spin.
\item $\Delta(p)$ is a projector: $\Delta(p)=\Delta(p)\Delta(p)$.
\end{itemize}
$I$ represents the total angular momentum while $s$, the spin, is only the intrinsic
rotation of the particle/field: $|s\ra$ is a vector of the $I$ representation i.e
$s\le I$. From the field point of view, $I=s$ represents the fundamental field $\phi^{(s)}$
of spin $s$, while the representation $I=s+k$ corresponds to the $k$-th derivative of the field
$\pp^k\phi^{(s)}$.

The unique solution to these constraints is:
\be
\Delta(p)=D^I\left(s(p)\right)|s\ra\la s| D^I\left(s(p)\right)^{-1}.
\ee
This is easy to generalize to different $I$'s on the left and right hand side. Now
$$
\Delta(p)=D^I\left(s(p)\right)|s\ra\la s| D^{I'}\left(s(p)\right)^{-1}
$$
corresponds to the correlation $\la \pp^k\phi^{(s)}\,\pp^{k'}\phi^{(s)}\ra$ with $I=s+k, I'=s+k'$.

For $I=s=0$, the scalar field, the propagator is obviously $\Delta=1$.
For the spin-$1/2$ particle, $I=s=1/2$, the definition of the section $s(p)$ reads in spinor
notations\footnotemark:
$$
D^{\f{1}{2}}(s(p))
\mat{cc}{1 & 0 \\ 0 & -1}
D^{\f{1}{2}}(s(p))^{-1}=
\f{{}^{(3)}\vec{p}}{m}\cdot\vec{\sigma}=
\f{1}{m}\mat{cc}{p^0 & p_+ \\ p_- & -p^0},
$$
where the $\sigma_i$'s are the Pauli matrices normalized to $\sigma_i\sigma_i=Id$.
So the propagator for $s=+1/2$ is:
\bes
\Delta_{+1/2}(p)&=&
D^{\f{1}{2}}(s(p))
\mat{cc}{1 & 0 \\ 0 & 0}
D^{\f{1}{2}}(s(p))^{-1}\\
&=&
D^{\f{1}{2}}(s(p))
\,\f{1}{2}\left(
Id+
\mat{cc}{1 & 0 \\ 0 & -1}
\right)\,
D^{\f{1}{2}}(s(p))^{-1} =
\f{1}{2m}\left(m+\vec{p}\cdot\vec{\sigma}\right),
\ees
and similarly for the negative helicity $s=-1/2$:
\be
\Delta_{+1/2}(p)=\f{1}{2m}\left(m-\vec{p}\cdot\vec{\sigma}\right).
\ee
And we recognize the usual expressions for a spinor field for the two helicities.

\footnotetext{The spinor representation of a 3-vector $(u_0,u_x,u_y)$ is:
$$
\mat{cc}{u_0 & u_+=u_x+iu_y \\ u_-=u_x-iu_y & -u_0}.
$$}

For a $I=1$ field, the section is defined by:
$$
D^{I=1}(s(p)).\mat{c}{1 \\0 \\0}=\f{1}{m}\mat{c}{p_0\\p_x\\p_y}
\quad\Rightarrow\quad
D(s(p))=R(\what{p})B(|{}^2\vec{p}|)R(\what{p})^{-1},
$$
with
$$
R(\what{p})=\mat{ccc}{1&0&0\\0&\f{p_x}{|{}^2\vec{p}|}&-\f{p_y}{|\vec{p}|}\\
0&\f{p_y}{|\vec{p}|}&\f{p_x}{|\vec{p}|}},
\qquad
B(|{}^2\vec{p}|)=\mat{ccc}{\f{p_0}{m}&\f{|\vec{p}|}{m}&0\\-\f{|\vec{p}|}{m}&\f{p_0}{m}&0\\0&0&1}.
$$
If we look at the spin $s=0$ field, the projector is:
$$
|I=1,0\ra\la0|=\mat{ccc}{1&0&0\\0&0&0\\0&0&0},
$$
so that the propagator is
$$
\Delta^{I=1,s=0}_{\mu\nu}(p)=\f{p_\mu p_\nu}{m^2},
$$
which naturally corresponds to the correlation $\la\pp_\mu\phi \pp_\nu\phi\ra$
of a scalar field $\phi$.
To recover the usual propagator for a spin-1 field, $s=1$,
we in fact sum over the states $|s=-1\ra$ and $|s=+1\ra$. The projector then reads:
$$
|-\ra\la-|+|+\ra\la+|=
\mat{ccc}{0&0&0\\0&1&0\\0&0&1},
$$
so that the propagator is simply:
\be
\Delta^{I=s=1}_{\mu\nu}(p)=\delta_{\mu\nu}-\f{p_\mu p_\nu}{m^2}.
\ee

\subsection{Inserting spinning particles in the spin foam}

Now, let us look at the Feynman amplitude of a graph with massive and spinning particles.
The part concerning the mass will not change: we still impose the constraint
$\delta(|\vec{p}|-m)$ on each edge and conservation of the momentum at each vertex.
Moreover we now need to take into account the further degrees of freedom corresponding
to the spin (or intrinsic rotation) of the particles. On each edge $e$, we have a
spin $s_e$ and we put the corresponding propagator $\Delta^{I=s_e}(p_e)$. At the vertices,
we need to take into account the interaction between spinning particles: we include
$SU(2)$ intertwiners intertwining between the representations $I=s_e$ of the edges meeting
at the vertices. On the whole, the Feynman evaluation reads:
\be
I_\Gamma=\int\prod_ed^3\vec{p_e}\,
\prod_e\f{\delta(|p_e|-m_e)}{4\pi m_e}\Delta^{I_e=s_e}_{l_el'_e}(p_e)\,\times\,
\prod_v\delta\left(\sum_{e|v=t(e)}\vec{p}_e-\sum_{e|v=s(e)}\vec{p}_e\right)
C^{I\dots}_{l\dots}.
\ee
Let us point out that we need not in principle take $I_e=s_e$, then the propagators
would live in the representations $I_e$  and the intertwiners $C$ intertwine between
the $I_e$'s: the spins $s_e$ would only enter as projectors in the definition
of the propagators $\Delta_{ll'}(p)=D_{ls}(s(p))D_{sl'}(s(p))^{-1}$.

\medskip

It is straightforward to write
this amplitude in its dual form solving the constraints imposed by momentum
conservation at the vertices. Indeed, adding spins doesn't modify anything, so that
the amplitude corresponding to the graph embedded in the 3 triangulation reads:
\be
I_\Gamma=
\int\prod_\eb d^3\vec{u}_\eb
\prod_{e\notin\Gamma}\delta(u_e)
\prod_{e\in\Gamma} \f{\delta(|u_e|-m_e)}{4\pi m_e}
\Delta^{I_e=s_e}_{l_el'_e}(p_e)\,\times\,C^{I\dots}_{l\dots}.
\ee
So given the graph $\Gamma$, we identify the plaquettes dual to the edge of
$\Gamma$. These form a tube around the graph. We evaluate the holonomy $p_e$
around each plaquette, then joining them all into a spin network -a graph drawn
on the surface of the tube- using the intertwiners $C^{I\dots}_{l\dots}$. Finally
the Feynman evaluation id the product of the mass shell conditions with the spin
network evaluation.

\medskip

Let us now transpose this formula to the non-abelian case in order to encode
spinning particles in the Ponzano-Regge model, such that its abelian
limit will reduce to the above Feynman amplitude.

The mass shell condition translates into $\delta_{\theta_e}(g_e)$, which imposes that
$g_e=uh_\theta u^{-1}$ for some $u\in\SU(2)/\U(1)$. $u$ represents the momentum of the
particle so that the spinning propagator translates to $D^{I=s}_{ls}(u)D^{I=s}_{sl'}(u^{-1})$.
Finally, we contract these propagators using the intertwiners $C$'s which defines
the physical content of the quantum field theory we would like to encode on the spin foam.
The final amplitude is:
\be
Z_{\Gamma,\{\theta_e,s_e\}}=
\int\prod_{\eb}dg_\eb\,
\prod_{e\notin\Gamma}\delta(g_e)
\prod_{e\in\Gamma}\f{V_H}{V_G\Delta(\theta_e)}
\int_{\SU(2)/\U(1)} du_e\delta(guh_\theta u^{-1})
D^{I=s_e}_{l_e s_e}(u)D^{I=s_e}_{s_e l'_e}(u^{-1})
\times C^{I\dots}_{l\dots}.
\ee


\end{document}